%% file: iacrdoc.tex
\documentclass[journal=tches,preprint]{iacrtrans}
\usepackage[utf8]{inputenc}

\usepackage{algorithm}
\usepackage{algpseudocode}
\usepackage{verbatim}
\usepackage{multirow}

\author{Ahmet Can Mert\inst{1} \and Aikata\inst{1} \and Sunmin Kwon\inst{2}  \and Youngsam Shin\inst{2} \and Donghoon Yoo\inst{2} \and Yongwoo Lee\inst{2} \and Sujoy Sinha Roy\inst{1}}
\institute{IAIK, Graz University of Technology, Graz, Austria \email{{ahmet.mert,aikata,sujoy.sinharoy}@iaik.tugraz.at} \and Samsung Advanced Institute of Technology, Suwon, Republic of Korea \email{{sunmin7.kwon,youngsam.shin,say.yoo,yw0803.lee}@samsung.com}}

\title[Medha: Microcoded Hardware Accelerator for Processing Encrypted Data]{Medha: \textit{M}icrocoded \textit{H}ardware \textit{A}ccelerator for computing on  \textit{E}ncrypted \textit{D}ata}

\begin{document}

\maketitle


\begin{abstract}

Homomorphic encryption enables computation on encrypted data, and hence it has a great potential in privacy-preserving outsourcing of computations to the cloud. Hardware acceleration of homomorphic encryption is crucial as software implementations are very slow. In this paper, we present design methodologies for building a programmable hardware accelerator for speeding up the cloud-side homomorphic evaluations on encrypted data.  

First, we propose a divide-and-conquer technique that enables homomorphic evaluations in the polynomial ring $R_{Q,2N} = \mathbb{Z}_{Q}[x]/(x^{2N} + 1)$ to use a hardware accelerator that has been built for the smaller ring $R_{Q,N} = \mathbb{Z}_{Q}[x]/(x^{N} + 1)$. The technique 
makes it possible to use a single hardware accelerator flexibly for supporting several homomorphic encryption parameter sets.

Next, we present several architectural design methods that we use to realize the flexible and instruction-set accelerator architecture, which we call `Medha'.
At every level of the implementation hierarchy, we explore possibilities for parallel processing. Starting from hardware-friendly parallel algorithms for the basic building blocks, we gradually build heavily parallel RNS polynomial arithmetic units. 
Next, many of these parallel units are interconnected elegantly so that their interconnections require the minimum number of nets, therefore making the overall architecture placement-friendly on the platform. As homomorphic encryption is computation- as well as data-centric, the speed of homomorphic evaluations depends greatly on the way the data variables are handled. For Medha, we take a memory-conservative design approach and get rid of any off-chip memory access during homomorphic evaluations.

Finally, we implement Medha in a Xilinx Alveo U250 FPGA and measure timing performances of the microcoded homomorphic addition, multiplication, key-switching, and rescaling routines for the leveled fully homomorphic encryption scheme RNS-HEAAN at 200 MHz clock frequency. For the large parameter sets $(\log Q, N) = (438, 2^{14})$ and $(546, 2^{15})$, Medha achieves accelerations by up to $68\times$ and $78\times$ times respectively compared to a highly optimized software implementation Microsoft SEAL running at 2.3 GHz. \footnote{Paper will appear at IACR Transactions on Cryptographic Hardware and Embedded Systems 2023.}

\end{abstract}

\input{1.intro}
\input{2.background}

\input{3.flex_technique}
\input{4.low_arith_units}
\input{5.high_arith_units}

\input{6.results}
\input{7.conclusion}
\section*{Acknowledgement}

This work was supported in part by the Samsung Electronics co. ltd., Samsung Advanced Institute of Technology and the State Government of Styria, Austria -- Department Zukunftsfonds Steiermark. We thank the anonymous reviewers for their useful suggestions and comments. 
We also thank Patrick Schaumont for shepherding the paper.

\bibliographystyle{alpha}
\bibliography{biblio}


\appendix
\section{Modular Arithmetic Architectures}\label{app:modred}
For implementing modular arithmetic, we use hardware-friendly algorithms and optimizations so that we can optimize these fundamental arithmetic blocks for both area and speed. {Modular addition and subtraction} are the simplest blocks and they are implemented using adder and subtractor circuits made of configurable fabric logic, i.e., LUTs.

For implementing \textit{modular multiplication}, we use the DSP units available in FPGAs so that the highest throughput can be achieved. 
We used several layers of pipeline registers to meet high clock-frequency constraints. One integer multiplier (60-bit or 54-bit) consumes 10 DSP slices.  
The most commonly used methods for performing modular reduction are based on the Barrett or Montgomery methods. The SEAL library~\cite{SEAL3.6} and the hardware architecture of the HEAX processor~\cite{heax} use the Barrett reduction technique for reducing the results of integer multiplications. Both Barrett and Montgomery methods are based on multiplications.
Another method for implementing the modular reduction operation is using a table-based modular reduction approach as proposed in~\cite{roy_tc2018}. In this method, the result of a multiplication is reduced in multiple steps where each step reduces a part near the most significant bit using a pre-computed look-up table.
Although this method does not use any multipliers, it increases LUT utilization on FPGA and does not provide an optimal solution when the modulus is constant.

When the modulus is selected as a sparse prime, the reduction operation can be performed efficiently using only add and shift operations as proposed in~\cite{zhang_ches21}.
In this work, we use reduction-friendly pseudo-Mersenne primes and employ the fast add-shift-based modular reduction method to save both DSPs and LUTs~\cite{zhang_ches21}. 
For example, the first modulus in the RNS-basis ($q_0$) is a 60-bit sparse prime $2^{59} + 2^{25} + 2^{22} - 2^{20} + 1$ and the result of a multiplication (120 bit) is reduced using the relation $2^{59} \equiv - 2^{25} - 2^{22} + 2^{20} - 1 \pmod{q_0}$ recursively.
This approach saves up to 45\% LUT units compared to the table-based modular reduction method.

\section{NTT Algorithm}\label{app:ntt}

\begin{algorithm}[H]
\small
\caption{Decimation-in-time (DIT) Forward NTT}
\textbf{In:} $\boldsymbol{a} \in R_{Q,N}$ and $\psi \in \mathbb{Z}_q$ ($2N$-th root of unity)\\
\textbf{Out:} $\tilde{\boldsymbol{a}} = \mathtt{NTT}(\boldsymbol{a}) \in R_{Q,N}$
\begin{algorithmic}[1]
	\State $\tilde{\boldsymbol{a}} \leftarrow \mathtt{BitReverse}(\boldsymbol{a})$ \Comment{Permutation of coefficients}
	\For{$m=2$ to $N$ by $m=2m$}
		\State $w \leftarrow 1$, $\psi_m \leftarrow \psi^{N/m}$ \;
		\For{$j=0$ to $m/2-1$}	\Comment{Butterfly loop}		
			\For{$k=0$ to $N-1$ by $m$}	 
				\State $u \leftarrow \tilde{\boldsymbol{a}}[k+j], t \leftarrow \tilde{\boldsymbol{a}}[k+j+m/2]$ 
				\State $\tilde{\boldsymbol{a}}[k+j] \leftarrow u + w \cdot t$ \;
				\State $\tilde{\boldsymbol{a}}[k+j+m/2] \leftarrow u - w \cdot t$ \;					 
			\EndFor
			\State $w \leftarrow w \cdot \psi_m$ \;
		\EndFor
	\EndFor
\end{algorithmic}\label{algo:dit_ntt}
\end{algorithm} 

\end{document}

%% file: 1.intro.tex
\section{Introduction} \label{sec:intro}

Cloud computing services are very popular and provide high-performance computational resources to  the users~\cite{armbrust2010view}. Despite its advantages, conventional cloud computing has security and privacy risks as the data of the user, becomes visible (as plaintext) during any computation in the cloud. Isolation techniques are followed with certain trust assumptions. Yet, in recent years several data leaks have been reported.

Fully Homomorphic Encryption (FHE)~\cite{RIVEST78} enables logical and arithmetic operations on encrypted data without requiring any decryption of the data. Clients can keep their data in encrypted format in the cloud and perform computations, e.g., inference, prediction, statistics, etc., directly on the encrypted data without revealing their confidential data. Similarly, the cloud-based service providers do not need to reveal their trade secrets, e.g., models or methods, to the clients. If a service provider keeps the model or method secret, a client cannot run the computation locally on her plaintext data. In this way, FHE solves conflicting privacy needs of users and service providers.

In 2009, Gentry constructed the first FHE scheme~\cite{GENTRY09}. FHE quickly gained interest from both academia and industry. During the last 10 years, faster and faster FHE schemes started appearing with orders of magnitude improvements in performance. There are several FHE or leveled FHE schemes in the literature. The difference between an FHE and a leveled-FHE is that the latter could perform computations correctly only up to a certain complexity level whereas the first one could do arbitrary computations. It is possible to transform a leveled-FHE into an FHE by introducing a special procedure `bootstrapping' that refreshes ciphertexts. This paper focuses on the hardware acceleration aspects of leveled-FHE. In the remaining part of the paper, we will use the term HE to represent leveled-FHE. For evaluating arithmetic operations homomorphically, BFV~\cite{DBLP:journals/iacr/FanV12} and BGV~\cite{DBLP:journals/eccc/BrakerskiGV11} are popular. TFHE~\cite{chillotti2020tfhe} is efficient for evaluating Boolean gates. For performing computations on encrypted \emph{real numbers}, HEAAN~\cite{ckks_scheme} and its Residue Number System (RNS) variant RNS-HEAAN~\cite{RNS_HEAAN18} are efficient. 
In fact, RNS-HEAAN is the fastest scheme for performing approximate computations on the encrypted \emph{real} data.

\subsection{Related hardware acceleration works and motivation} \label{sec:related_works}
Although a decade of research in algorithmic and mathematical optimizations has made HE schemes orders of magnitude faster than their first-generation counterparts, homomorphic evaluations in software are four to five orders of magnitude slower than equivalent computations on the plaintext. Therefore, the hardware acceleration of HE is crucial in reducing this performance gap. While it is true that developing an accelerator for homomorphic evaluation will take a longer design-time and more effort than writing a software program, the reduction in computation time and energy consumption in the cloud server will be huge in long-term.

Earliest reported FPGA-base accelerator~\cite{wang_fpga_gentry} and ASIC-based accelerator~\cite{wang_asic_gentry} targeted speeding up the 768K-bit modular multiplication of the first generation integer-based homomorphic encryption scheme~\cite{gentry_halevi_fhe}. In the following years several accelerator architectures~\cite{roy_ches2015, takeshita2020, Reis_2019, Reis_2020, feldmann_2021f1, xin_iscas21, BTS_isca, CraterLake} have implemented selected building blocks of homomorphic encryption on ASIC and FPGA platforms, or presented simulation-based performance estimates without making real prototypes. 
While such works indicate that ASIC and FPGA platforms have the potential to accelerate HE, they do not fully capture the engineering challenges that appear only in the actual hardware implementations, especially when the architectures are very large. Therefore, simulation-based performance estimates may change greatly when real hardware accelerators for homomorphic encryption are made.

Before we describe related hardware acceleration works for cloud-side homomorphic computing, we would like to mention that hardware acceleration could also be used to speedup client-side encryption and decryption operations. One example accelerator is~\cite{fhe_enc_hw}. Compared to homomorphic evaluation on encrypted data, homomorphic encryption and decryption are much simpler and less frequent. They are quite similar to simple ring-LWE encryption and decryption used in lattice-based post-quantum cryptography. In homomorphic encryption the real bottleneck is the slowness of cloud-side homomorphic evaluations. Therefore, in this paper we focus only on the hardware acceleration of cloud-side evaluations. Readers may also study symmetric-homomorphic hybrid protocols where a client simply encrypts the data using an homomorphic encryption friendly block-cipher such as Pasta~\cite{pasta} to save communication bandwidth. Thereafter, the cloud evaluates the expensive block-cipher decryption homomorphically before computing the actual task.

Recently, several papers~\cite{feldmann_2021f1,BTS_isca,CraterLake,ARK} proposed ASIC-based high-end accelerator architectures and claimed three to four orders of magnitude speedups with respect to software for performing homomorphic evaluations. These works use simulation and logic synthesis for obtaining performance and area estimates respectively, without going through the complete ASIC design flow or fabricating a real ASIC chip. 
Following the chip fabrication price estimates~\cite{fab_pricing}, fabricating these ASIC chips will require millions of dollars of investments.

To the best of our knowledge, CoFHEE~\cite{cofhee} is the only ASIC accelerator that has been fabricated and proven in silicon. CoFHEE's total die area is 15 mm$^2$ and it accelerates homomorphic evaluations only up to 2.5$\times$ compared to the SEAL software library. In their year-long effort to design the chip, the authors follow the complete ASIC design flow, implement a custom clock distribution network, perform pre-silicon verification using simulation and FPGA prototyping, and finally perform post-silicon validation to know that their ASIC chip works correctly. {The authors of CoFHEE raise concerns about the feasibility of F1~\cite{feldmann_2021f1} in silicon (see Sec.~9 of~\cite{cofhee}).}

Although FPGAs are slower than ASIC platforms, their relatively shorter design cycle, re-programmability to fix bugs easily, reusability, and significantly cheaper price make FPGAs popular for implementing performance-critical algorithms. 
The FPGA-based programmable accelerators~\cite{roy_hpca, he_aws} demonstrated latency reductions by one order compared to software implementations. `HEAX'~\cite{heax} obtained more than two orders of magnitude throughput with respect to software implementations using one Intel FPGA. While the speedup is impressive, a limitation of HEAX is that it is not programmable and its block-pipelined architecture was designed specifically for the key-switching of RNS-HEAAN. Contrary to HEAX, the programmable accelerator~\cite{roy_hpca} uses the same computational resources to execute several homomorphic evaluation routines. While programmability is a desired feature in accelerators, the one order speedup of HEAX~\cite{heax} over the programmable processor~\cite{roy_hpca, he_aws} may give an impression that block-pipelined and specifically optimized accelerators are significantly superior to flexible accelerators for HE. 

In this work, we dig deep into architectural explorations to see if programmable and flexible accelerator architectures can be built without sacrificing performance. The availability of a programmable accelerator will make it possible to run and accelerate several types of homomorphic evaluation routines without requiring a new accelerator architecture. 
However, developing a real prototype of a flexible and high-performance accelerator for homomorphic encryption is full of design challenges. This motivates us to see how far we can push homomorphic computing on encrypted data in practice using programmable hardware. Sharing the experiences and methodologies for designing a real high-performance accelerator will help the research community identify the actual engineering challenges as well as future research directions for potential performance improvements. 

Another research gap is the lack of a parameter-flexible accelerator. Homomorphic applications of different complexities (multiplicative depths) demand different parameter sets. Hence, supporting several parameter sets is another important yet currently unfulfilled requirement for the cloud-side accelerators. Almost all of the reported accelerators~\cite{roy_tc2018,roy_hpca,heax,he_aws} have been designed for specific parameter sets and they lack the flexibility to support more than one parameter set. That motivates us to design a programmable and parameter-flexible accelerator for homomorphic encryption.

\subsection{Contributions}

To address the above-mentioned research gaps, we design a programmable and parameter-flexible hardware accelerator architecture `Medha' and implement it in a Xilinx Alveo U250 Card. Medha accelerates the homomorphic addition, multiplication, key-switching, re-scaling, and rotation operations of the RNS-HEAAN~\cite{RNS_HEAAN18} scheme for large parameter sets by around two orders of magnitude compared to software implementations. 

The main contributions of our paper reside at both algorithmic and architecture levels. First, we propose a design methodology that offers the flexibility to support several polynomial degrees using a fixed hardware accelerator. It efficiently performs homomorphic computations on ‘large-degree’ ciphertext-polynomials using compute units that have been optimized to handle ‘small-degree’ ciphertext-polynomials. Therefore, several homomorphic applications can be accelerated using the same architecture.
Additionally, the proposed methodology reduces the on-chip memory and logic requirements.

On the architecture side, our main contributions are in the high-levels of the implementation hierarchy where different compute and memory elements are organized. To compute the arithmetic of residue polynomials, we design a novel Residue Polynomial Arithmetic Unit (RPAU) pragmatically. Our RPAU contains a multi-core Number Theoretic Transform (NTT) unit for polynomial multiplication, two parallel sets of dyadic arithmetic units, and a customized on-chip memory for storing operand and resultant residue polynomials. 
The designed RPAU is an instruction-set architecture. We can execute dyadic arithmetic and NTT instructions in parallel. This parallelism is very useful in minimizing the cycle count of key-switching operation, which is the costliest subroutine in HE. We observe around 40\% reduction in the latency at the cost of around 20\% increase in the area.

A memory-conservative design approach is followed to save on-chip memory elements for useful computations. A customized on-chip memory is designed to store residue polynomials inside the RPAU. Even for the largest supported polynomial degree with $2^{15}$ coefficients, the on-chip memory is able to store all the residue polynomials during a homomorphic multiplication and key-switching, therefore eliminating the need for any off-chip data exchange during a computation (which is very slow). We are the first to report fully on-chip computation of the two HE subroutines for such large-degree polynomials. 

At the highest level of the implementation hierarchy, several RPAUs are instantiated and interconnected. The parallel RPAUs must perform data exchanges between themselves during the modulus switching steps of the key-switching and rescaling operations. Trivially connecting every RPAU to the remaining RPAUs demands a quadratic number of nets and makes actual implementation infeasible when there are several large RPAUs in the architecture. Therefore, finding an optimal way of interconnecting the RPAUs is critical due to two main reasons. Firstly, because each RPAU consumes a large area and has thousands of bits of input/output ports, their placement on the design platform becomes a challenging engineering problem. Previous works, e.g.,~\cite{roy_ches2015, roy_tc2018} bypassed that engineering problem by performing data exchanges happen via a shared off-chip memory at a great performance cost. Secondly, if the data transfer rate is compromised to reduce the number of interconnects, then there will be a drastic impact on the performance of key-switching and rescaling. After studying different ways of interconnecting the RPAUs, we propose a `ring' styled interconnection with efficient scheduling of data transfers. The ring reduces the number of interconnects to a linear complexity without introducing any performance loss. We observe that the proposed ring is crucial for making large HE implementations feasible on SLR-based very large Xilinx FPGAs. 
    
Besides the above-mentioned main contributions, we make optimizations at the lower levels of the implementation hierarchy where polynomial and coefficient operations are performed. We implement a unified and multi-core NTT-based multiplier with optimal scheduling for memory reads and writes. We use hardware-friendly parameters (e.g., word-size, primes, etc.) and parallel algorithms to perform fast modular arithmetic. \\

\noindent\textbf{Organization:}
The paper is organized as follows. Sec.~\ref{sec:background} presents a brief background. Next, we present the proposed flexible design methodology for the polynomial degree and its applications to the RNS-HEAAN scheme in Sec.~\ref{sec:flex}.
In Sec.~\ref{sec:medha}, the proposed flexible instruction-set accelerator is presented. 
The accelerator architecture is realized hierarchically starting from low-level polynomial arithmetic units in Sec.~\ref{sec:imp_low_level} and then organizing different compute and memory elements in Sec.~\ref{sec:imp_rpau}. 
Detailed experimental results and comparisons are provided in Sec.~\ref{sec:results} and the final section draws the conclusions.


%% file: 2.background.tex
\section{Background} \label{sec:background}

\subsection{Notation}\label{sec:background_notation}

Let $\mathbb{Z}_Q$ represent the ring integers in the range $[0,Q-1]$. Any modular reduction by a modulus $Q$ in $[0,Q-1]$ is denoted as $[.]_q$. 
The polynomial ring $R_{Q,N}=\mathbb{Z}_Q[x]/(x^N+1)$ contains polynomials of degree at most $N-1$ with coefficients in $\mathbb{Z}_Q$.
An integer is represented using a normal font and lowercase letter, e.g., $a \in \mathbb{Z}_Q$. A polynomial is represented using a bold font and lowercase letter, $\boldsymbol{a} \in R_{Q,N}$.
%
%
When a residue number system (RNS) is used with a composite modulus $Q = \prod_{i=0}^{L-1} q_i$, 
a polynomial in $R_{Q,N}$ becomes a vector of residue polynomials in the RNS.  
%
Let $\boldsymbol{a}[i] \in R_{q_i,N}$ be the $i$-th residue polynomial in the RNS representation of $\boldsymbol{a} \in R_{Q,N}$.
%
$R_{Q,N}^k$ represents a $k$-tuple of polynomials from $R_{Q,N}$. 
Thereafter, the $i$-th residue polynomial of the $j$-th member from the tuple $\boldsymbol{a} \in R_{Q,N}^k$ is represented as $\boldsymbol{a}[j][i] \in R_{q_i,N}$.

The Number Theoretic Transform (NTT) of a polynomial $\boldsymbol{a}$ is represented by $\tilde{\boldsymbol{a}}$. The multiplication between two elements of a ring is denoted by the $\cdot$ operator. The coefficient-wise multiplication between two polynomials is denoted by the $\star$ operator. Multiplying all the coefficients of a polynomial $\boldsymbol{a}$ by an integer scalar $c$ is denoted by $\boldsymbol{a} \odot c$ or $c \odot \boldsymbol{a}$.

\subsection{Homomorphic Encryption}\label{sec:background_he}

In a typical homomorphic encryption protocol, there are two parties: a client and a cloud server. The cloud contains data encrypted (i.e., ciphertext) by the client, and the client performs computations on its encrypted data in the cloud. At the end of computations, the client receives the encrypted results from the cloud and performs decryptions locally to recover the plaintext results. 
The encryption and decryption operations are performed using the secret-key and private-key of the client, respectively.

Several ideal lattice-based homomorphic encryption schemes, e.g., BGV~\cite{DBLP:journals/eccc/BrakerskiGV11}, BFV~\cite{DBLP:journals/iacr/FanV12}, and HEAAN~\cite{RNS_HEAAN18} use the following framework. 
%
%
Let, a client's secret-key be $\mathtt{sk} = (1, \boldsymbol{s}) \in R^2_{Q,N}$ and the corresponding public-key be $\mathtt{pk} = (\boldsymbol{b}, \boldsymbol{a}) \in R^2_{Q,N}$. Each key is a pair of polynomials in the polynomial ring $R_{Q,N}$ where $Q$ is the coefficient-modulus and $N$ is the polynomial ring degree. Client encrypts a message $\boldsymbol{m}$ using $\mathtt{pk}$ and obtains the ciphertext $\mathtt{ct} \leftarrow ( \boldsymbol{c}_0 = \boldsymbol{r} \cdot \boldsymbol{b} + \boldsymbol{e}_0 + \boldsymbol{m}, \boldsymbol{c}_1 = \boldsymbol{r}\cdot \boldsymbol{a} + \boldsymbol{e}_1) \in R^2_{Q,N}$ where $\boldsymbol{e}_i$ is a Gaussian distributed error-polynomial and $\boldsymbol{r}$ is a uniformly random polynomial. 
Let, a cloud contains two ciphertexts $\mathtt{ct}=(\boldsymbol{c}_0, \boldsymbol{c}_1)$ and $\mathtt{ct}'=(\boldsymbol{c}'_0, \boldsymbol{c}'_1) \in R^2_{Q,N}$ of the client as the encrypted messages $\boldsymbol{m}$ and $\boldsymbol{m}'$ respectively. The cloud can compute a valid encryption of $\boldsymbol{m}+\boldsymbol{m}'$ simply by adding the two ciphertexts as $\mathtt{ct}_\text{add} \leftarrow (\boldsymbol{c}_0+\boldsymbol{c}'_0, \boldsymbol{c}_1+\boldsymbol{c}'_1) \in R^2_{Q,N}$.
Computing an encryption of $\boldsymbol{m} \cdot \boldsymbol{m}'$ is relatively complex, scheme specific and involves several steps. First, the two ciphertexts are multiplied to obtain $\mathtt{ct}_\text{mult} = (\boldsymbol{c}_0\cdot \boldsymbol{c}'_0, \boldsymbol{c}_0\cdot \boldsymbol{c}'_1 + \boldsymbol{c}_1 \cdot \boldsymbol{c}'_0, \boldsymbol{c}_1 \cdot \boldsymbol{c}'_1) \in R^3_{Q,N}$. This intermediate result has three polynomial components and could be decrypted using $(1, \boldsymbol{s}, \boldsymbol{s}^2)$ but not using $\mathtt{sk}=(1,\boldsymbol{s})$. Next, a special operation known as the `Key-Switching', is used to transform the three-component ciphertext $\mathtt{ct}_\text{mult}$, which is decryptable under $(1, \boldsymbol{s}, \boldsymbol{s}^2)$, into the two-component ciphertext $\mathtt{ct}_\text{relin}$ decryptable under $(1, \boldsymbol{s})$. In this context, key-switching is called re-linearization as it produces a linear ciphertext.

\begin{figure}[!t]
\centering
\includegraphics[width=3.9in]{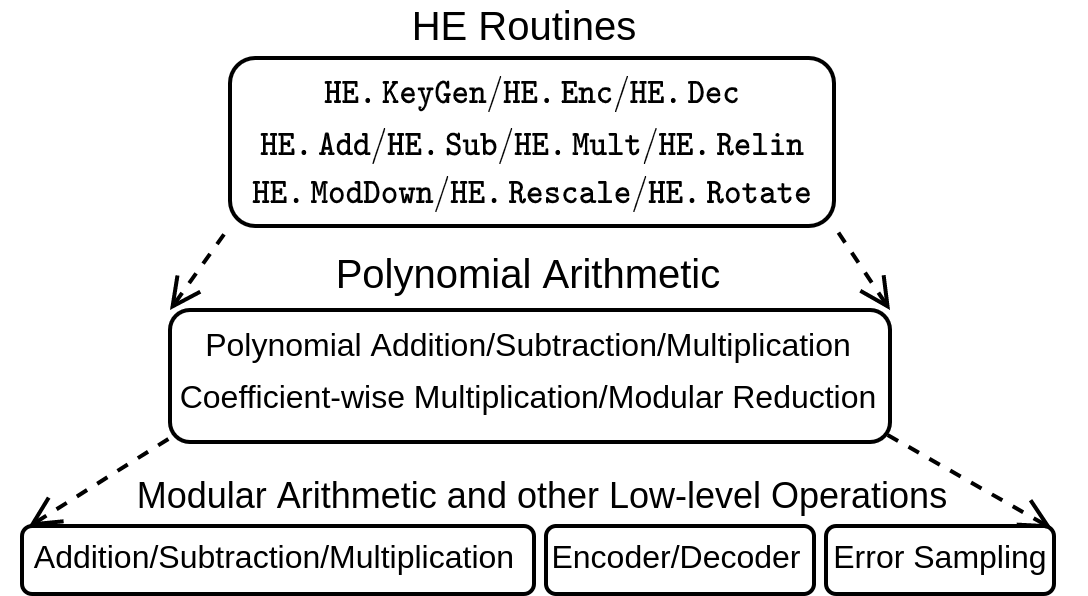}
\caption{Implementation hierarchy of homomorphic encryption. Key generation, encryption, and decryption are performed on the user side. These user-side operations also include encoding, decoding, and error sampling at the lowest level. Homomorphic evaluations on ciphertexts using homomorphic addition, subtraction, multiplication, relinearization, etc., are performed at the cloud side. Our hardware accelerator is designed for accelerating cloud-side operations which are significantly more expensive than user-side operations.}
\label{fig:he_hieararchy}
\end{figure}

Fig.~\ref{fig:he_hieararchy} shows the hierarchy of different operations that are used in a homomorphic application. At the highest level of this hierarchy, there are homomorphic procedures for performing computations (e.g., addition, multiplication, key-switching, etc.) on the ciphertexts. These high-level operations translate into the arithmetic of polynomials: polynomial addition, polynomial subtraction, polynomial multiplication, coefficient-wise multiplication, coefficient-wise modular reduction, and coefficient-wise scalar multiplication. Finally, the lowest level of this hierarchy is composed of modular arithmetic.

\subsection{Residue Number System (RNS)}\label{sec:background_rns}

In HE, typically the size of $Q$ is several hundreds to several thousands of bits. Arithmetic operations with a large coefficient modulus require multi-precision arithmetic and it reduces the performance of HE-based applications. The Residue Number System (RNS) is a popular choice for implementing HE on hardware and software platforms as it enables the parallelization of computations. To work with the RNS, the modulus $Q$ is chosen to be a product of small coprimes $q_i$ such that $Q = \prod_{i=0}^{L-1} q_i$. These coprimes form the base $\mathcal{B} = \{q_0, \ldots, q_{L-1}\}$ of the RNS. Any long integer $a \in \mathbb{Z}_Q$ gets mapped into the set of small residues $a_i = a \pmod{q_i}$ in the RNS. On the other hand, a set of small residues $a_i$ in the RNS can be combined together using the Chinese Remainder Theorem (CRT) to obtain the integer $a \in \mathbb{Z}_Q$. 

In ideal lattice-based homomorphic encryption schemes, with the application of RNS, a polynomial $\boldsymbol{a} \in R_{Q,N}$ gets mapped into the set of residue polynomials in the RNS base. These residue polynomials have small coefficients and therefore they can be processed without requiring expensive multi-precision arithmetic operations. Additionally, on platforms with parallelism, e.g., hardware platforms or superscalar processors, a set of residue polynomials can be processed in parallel. In summary, the application of RNS makes implementations of homomorphic encryption simpler, modular, and faster.

\subsection{RNS-HEAAN}\label{sec:background_heaan}
RNS-HEAAN~\cite{RNS_HEAAN18} is an RNS variant of the original HEAAN~\cite{ckks_scheme} homomorphic encryption scheme. The original HEAAN~\cite{ckks_scheme} scheme uses a large modulus (which is a power of two) and its expensive multi-precision integer arithmetic becomes a performance bottleneck. Our hardware accelerator Medha has been optimized to accelerate the cloud-side homomorphic operations of RNS-HEAAN. 
In this section, we briefly describe the cloud side arithmetic routines of the RNS-HEAAN variant that is used in SEAL~\cite{SEAL3.6}. 
We chose SEAL's RNS-HEAAN variant for our implementation to make fair performance comparisons with SEAL~\cite{SEAL3.6} and HEAX~\cite{heax}.

To use RNS-HEAAN in an application, the first step is to set up the scheme parameters such as the degree $N$ of the polynomial ring, size of the maximum ciphertext modulus $Q_L$, RNS-base $\mathcal{B}_L$ with $L$ small prime moduli, etc., depending on the maximum multiplicative depth $L$ required by the application for a desired level of security. 
The multiplicative depth is the maximum number of consecutive homomorphic multiplications that can be performed before a wrong result is produced due to too much noise.  
%
After the setup phase, the key generation phase generates the public-private key pair and the relinearization key. The relinearization key which is a key-switching key, say $\mathtt{KSK} = (\mathtt{KSK}_0, \mathtt{KSK}_1)$, is sent to the cloud as it will be used to relinearize the result of a homomorphic multiplication using the key-switching procedure. 
Each of $\mathtt{KSK}_0$ and $\mathtt{KSK}_1$ is an $L$-tuple and resides in $R_{pQ_L, N}^L$. Here $p$ is a special prime that is used only during the key-switching operation. Sometimes we denote this special prime by $q_L$ and consider the extended RNS base to be $\mathcal{B}_L = \{q_0, \ldots, q_{L-1}, p\}$.
We use the notation $\mathtt{KSK}_{0}[i]$ to select the $i$-th element from the $L$-tuple. This element is in $R_{pQ_L, N}$ and therefore it consists of $(L+1)$ residue polynomials in the RNS representation. We use the notation $\mathtt{KSK}_{0}[i][j]$ to represent the $j$-th residue polynomial of $\mathtt{KSK}_{0}[i]$. In summary, each of $\mathtt{KSK}_0$ and $\mathtt{KSK}_1$ is a vector of $L(L+1)$ residue polynomials.
Similarly, the slot rotation operation requires a key-switching which uses Galois keys, $\mathtt{GK} = (\mathtt{GK}_0, \mathtt{GK}_1)$. The Galois keys are also sent to the cloud.
A ciphertext at the multiplicative level $l \leq L$ is represented using the RNS base $\mathcal{B}_{l} = \{q_0, \ldots, q_{l-1}\}$. 

To get a detailed description of RNS-HEAAN, the readers may follow the original publication~\cite{RNS_HEAAN18} and SEAL's~\cite{SEAL3.6} RNS-HEAAN implementation documentation. 
\newline

\noindent\textbf{RNS-HEAAN subroutines used in the Cloud:}
In the following part, we use the notation $Q_l$ to represent the ciphertext-modulus at the multiplicative level $l$ where $Q_l = \prod ^{l-1}_{i=0} q_i$ with $l < L$. It implicitly performs all arithmetic operations on the residue polynomials. From now on, we use just HEAAN to represent RNS-HEAAN. 
\begin{itemize}
    
    \item $\mathtt{HEAAN.Add}(\mathtt{ct}, \mathtt{ct}')$: It adds the respective polynomials of the two ciphertexts $\mathtt{ct} = ({\boldsymbol{c}_0,\boldsymbol{c}_1}) \in R^2_{Q_l, N}$ and $\mathtt{ct}' = ({\boldsymbol{c}'_0,\boldsymbol{c}'_1}) \in R^2_{Q_l, N}$, and computes $\mathtt{ct}_{\text{add}} = (\boldsymbol{d}_0, \boldsymbol{d}_1)$ where $\boldsymbol{d}_0 = \boldsymbol{c}_0 + \boldsymbol{c}'_0 \in R^2_{Q_l, N}$ and $\boldsymbol{d}_1 = \boldsymbol{c}_1 + \boldsymbol{c}'_1 \in R^2_{Q_l, N}$.
    
    \item $\mathtt{HEAAN.Mult}(\mathtt{ct}, \mathtt{ct}')$: It multiplies two input ciphertexts $\boldsymbol{c} = ({\boldsymbol{c}_0,\boldsymbol{c}_1}) \in R^2_{Q_l, N}$ and $\mathtt{ct}' = ({\boldsymbol{c}'_0,\boldsymbol{c}'_1}) \in R^2_{Q_l, N}$, and computes ${\boldsymbol{d}_0} = {\boldsymbol{c}_0\cdot \boldsymbol{c}'_0} \in R_{Q_l, N}$, ${\boldsymbol{d}_1} = {\boldsymbol{c}_0\cdot \boldsymbol{c}'_1} + {\boldsymbol{c}_1 \cdot \boldsymbol{c}'_0}  \in R_{Q_l, N}$, and ${\boldsymbol{d}_2} = {\boldsymbol{c}_1 \cdot \boldsymbol{c}'_1} \in R_{Q_l, N}$. The output is the non-linear ciphertext $\mathtt{dt}=(\boldsymbol{d}_0, \boldsymbol{d}_1, \boldsymbol{d}_2) \in R^3_{Q_l, N}$.
    
    \item $\mathtt{HEAAN.Relin}(\mathtt{dt},\mathtt{{KSK}})$: It re-linearizes the result of $\mathtt{HEAAN.Mult}$ and produces a ciphertext with two polynomial components so that it is decryptable under the secret key. Let $\boldsymbol{d}'_{2}[i] = \big[{\boldsymbol{d}_2}\big]_{q_i}$ for $0 \leq i < l$. Now compute $\mathtt{ct}''=(\boldsymbol{c}_0'', \boldsymbol{c}_1'')$ where ${\boldsymbol{c}_0''} = \sum^{l-1}_{i=0} \boldsymbol{d}'_{2}[i]\cdot \mathtt{{KSK}}_{0}[i] \in R_{pQ_l, N}$ and ${\boldsymbol{c}_1''} = \sum^{l-1}_{i=0} \boldsymbol{d}'_{2}[i] \cdot \mathtt{{KSK}}_{1}[i] \in R_{pQ_l, N}$. Finally, output the re-linearized ciphertext $\mathtt{ct}_\text{relin} = ({\boldsymbol{d}_0, \boldsymbol{d}_1}) + (\mathtt{HEAAN.ModDown}({\boldsymbol{c}_0''}),\mathtt{HEAAN.ModDown}({\boldsymbol{c}_1''})) \pmod {Q_l}$. The $\mathtt{HEAAN.ModDown}$ operation is used to reduce coefficient modulus from ${pQ_l}$ to ${Q_l}$ and is computationally similar to the rescaling operation. 
    
\end{itemize}

\noindent\begin{minipage}[t]{0.49\textwidth}
    \vspace{0pt}
    \begin{algorithm}[H]
    \renewcommand{\algorithmicrequire}{\textbf{In:}}
    \renewcommand{\algorithmicensure}{\textbf{Out:}}
    \caption{$\mathtt{HEAAN.Add}$ Algorithm}
    \begin{algorithmic}[1]
        \Require $\mathtt{ct}=(\tilde{\boldsymbol{c}}_{0}, \tilde{\boldsymbol{c}}_{1})$, $\mathtt{ct}'=(\tilde{\boldsymbol{c}}'_{0}, \tilde{\boldsymbol{c}}'_{1}) \in R_{Q_l,N}^2$

        \Ensure  $\mathtt{d}=(\tilde{\boldsymbol{d}}_{0}, \tilde{\boldsymbol{d}}_{1}) \in R_{Q_l,N}^{2}$
      
    	\State $\tilde{\boldsymbol{d}}_{0} \leftarrow \tilde{\boldsymbol{c}}_{0} + \tilde{\boldsymbol{c}}'_{0}$
    	\State $\tilde{\boldsymbol{d}}_{1} \leftarrow \tilde{\boldsymbol{c}}_{1} + \tilde{\boldsymbol{c}}'_{1}$
    \end{algorithmic}\label{algo:add_14}
    \end{algorithm}
\end{minipage}
\hfill
\begin{minipage}[t]{0.49\textwidth}
    \vspace{0pt}
    \begin{algorithm}[H]
    \renewcommand{\algorithmicrequire}{\textbf{In:}}
    \renewcommand{\algorithmicensure}{\textbf{Out:}}
    \caption{$\mathtt{HEAAN.Mult}$ Algorithm}
    \begin{algorithmic}[1]
        \Require $\mathtt{ct}=(\tilde{\boldsymbol{c}}_{0}, \tilde{\boldsymbol{c}}_{1})$, $\mathtt{ct}'=(\tilde{\boldsymbol{c}}'_{0}, \tilde{\boldsymbol{c}}'_{1}) \in R_{Q_l,N}^2$

        \Ensure  $\mathtt{d}=(\tilde{\boldsymbol{d}}_{0}, \tilde{\boldsymbol{d}}_{1}, \tilde{\boldsymbol{d}}_{2}) \in R_{Q_l,N}^{3}$
      
    	\State $\tilde{\boldsymbol{d}}_{0} \leftarrow \tilde{\boldsymbol{c}}_{0} \star \tilde{\boldsymbol{c}}'_{0}$, $\tilde{\boldsymbol{d}}_{2} \leftarrow \tilde{\boldsymbol{c}}_{1} \star \tilde{\boldsymbol{c}}'_{1}$
    	\State $\tilde{\boldsymbol{d}}_{1} \leftarrow \tilde{\boldsymbol{c}}_{0} \star \tilde{\boldsymbol{c}}'_{1} + \tilde{\boldsymbol{c}}_{1} \star \tilde{\boldsymbol{c}}'_{0}$
    \end{algorithmic}\label{algo:mult_14}
    \end{algorithm}
\end{minipage}

\noindent\begin{minipage}[t]{0.49\textwidth}
    \vspace{0pt}
    \begin{algorithm}[H]
    \renewcommand{\algorithmicrequire}{\textbf{In:}}
    \renewcommand{\algorithmicensure}{\textbf{Out:}}
    \caption{$\mathtt{HEAAN.ModDown}$ Algorithm}
    \begin{algorithmic}[1]
        \Require  $\tilde{\boldsymbol{d}} \in R_{pQ_l,N}$
        \Ensure   $\tilde{\boldsymbol{d}}' \in R_{Q_l,N}$
        
        \State $\boldsymbol{t} \leftarrow \mathtt{INTT}(\tilde{\boldsymbol{d}}[l])$ 
        
        \For{$i=0$ to $l-1$}	
    	
    	\State $\tilde{\boldsymbol{t}} \leftarrow \mathtt{NTT}(\big[\boldsymbol{t}\big]_{q_i})$ \Comment{in $\mathbb{Z}_{q_i}$}
    	\State $\tilde{\boldsymbol{d}'}[i] \leftarrow \big[p^{-1} \odot (\tilde{\boldsymbol{d}}[i] - \tilde{\boldsymbol{t}})\big]_{q_i}$
    
        \EndFor
    
    \end{algorithmic}\label{algo:moddown_14}
    \end{algorithm} 
\end{minipage}
\hfill
\begin{minipage}[t]{0.49\textwidth}
    \vspace{0pt}
    \begin{algorithm}[H]
\renewcommand{\algorithmicrequire}{\textbf{In:}}
\renewcommand{\algorithmicensure}{\textbf{Out:}}
\caption{$\mathtt{HEAAN.Rescale}$ Algorithm}
\begin{algorithmic}[1]
    \Require  $\tilde{\boldsymbol{d}} \in R_{Q_l,N}$
    \Ensure   $\tilde{\boldsymbol{d}}' \in R_{Q_{l-1,N}}$
    
    \State $\boldsymbol{t} \leftarrow \mathtt{INTT}(\tilde{\boldsymbol{d}}[l-1])$
    
    \For{$i=0$ to $l-2$}	
	
	\State $\tilde{\boldsymbol{t}} \leftarrow \mathtt{NTT}(\big[\boldsymbol{t}\big]_{q_i})$ \Comment{in $\mathbb{Z}_{q_i}$}
	\State $\tilde{\boldsymbol{d}'}[i] \leftarrow \big[q^{-1}_l \odot (\tilde{\boldsymbol{d}}[i] - \tilde{\boldsymbol{t}})\big]_{q_i}$

    \EndFor

\end{algorithmic}\label{algo:rescale_14}
\end{algorithm} 
\end{minipage}

\begin{algorithm}[H]
\renewcommand{\algorithmicrequire}{\textbf{In:}}
\renewcommand{\algorithmicensure}{\textbf{Out:}}
\caption{$\mathtt{HEAAN.Relin}$ Algorithm}
\begin{algorithmic}[1]
    \Require  $\mathtt{d} = (\tilde{\boldsymbol{d}}_{0}, \tilde{\boldsymbol{d}}_{1}, \tilde{\boldsymbol{d}}_{2}) \in R_{Q_l,N}^{3}$, $\tilde{\mathtt{KSK}}_0 \in R_{pQ_l,N}^l, \tilde{\mathtt{KSK}}_1 \in R_{pQ_l,N}^l$
    \Ensure   $\mathtt{d}' = (\tilde{\boldsymbol{d}}'_{0}, \tilde{\boldsymbol{d}}'_{1}) \in R_{Q_l,N}^{2}$
    
    \For{$j=0$ to $l-1$}	
	
	\State $\boldsymbol{d}_{2}[j] \leftarrow \mathtt{INTT}(\tilde{\boldsymbol{d}}_{2}[j])$ \Comment{in $\mathbb{Z}_{q_j}$}

    \EndFor

    \For{$j=0$ to $l$}	\Comment{Here $q_{l}$ is used to represent special prime $p$}	
	    \State $(\tilde{\boldsymbol{c}}''_{0}[j], \tilde{\boldsymbol{c}}''_{1}[j]) \leftarrow 0$

	\For{$i=0$ to $l-1$}

    	\State $\tilde{\boldsymbol{r}} \leftarrow \mathtt{NTT}(\big[\boldsymbol{d}_{2}[i]\big]_{q_j})$ \Comment{in $\mathbb{Z}_{q_j}$}

    	\State $\tilde{\boldsymbol{c}}''_{0}[j] \leftarrow \big[\tilde{\boldsymbol{c}}''_{0}[j] + \tilde{\mathtt{KSK}}_{0}[i][j] \star \tilde{\boldsymbol{r}}\big]_{q_j}$, $\tilde{\boldsymbol{c}}''_{1}[j] \leftarrow \big[\tilde{\boldsymbol{c}}''_{1}[j] + \tilde{\mathtt{KSK}}_{1}[i][j] \star \tilde{\boldsymbol{r}}\big]_{q_j}$

	\EndFor	
	
	\EndFor	

    \State $\tilde{\boldsymbol{d}}'_{0} \leftarrow \tilde{\boldsymbol{d}}_{0} + \mathtt{HEAAN.ModDown}(\tilde{\boldsymbol{c}}''_{0})$, $\tilde{\boldsymbol{d}}'_{1} \leftarrow \tilde{\boldsymbol{d}}_{1} + \mathtt{HEAAN.ModDown}(\tilde{\boldsymbol{c}}''_{1})$

\end{algorithmic}\label{algo:relin_14}
\end{algorithm} 

\begin{itemize}    
    \item $\mathtt{HEAAN.Rescale}(\boldsymbol{c})$: It takes a ciphertext-polynomial $\boldsymbol{c} \in R_{Q_l,N}$ with level $l$ and produces a ciphertext element with level $l-1$. Let $\boldsymbol{c}' \in R_{Q_{l-1},N}$ such that $\boldsymbol{c}'[i] = {\boldsymbol{c}[l]} \pmod{q_i}$ for $0 \leq i \leq l-1$. Then, compute $\boldsymbol{c}'' = \boldsymbol{c} - \boldsymbol{c}' \in R_{Q_{l-1},N}$. Finally, output the rescaled ciphertext element $\boldsymbol{c}'' =  q^{-1}_l \odot \boldsymbol{c}''  \in R_{Q_{l-1},N}$. To rescale a ciphertext that consists of two polynomials, the above procedure is applied to both the polynomials.
    
    \item\textbf{$\mathtt{HEAAN.Rotate}(\mathtt{ct},\mathtt{{GK}})$:}
    The slot rotation operation takes a ciphertext $\mathtt{ct} = (\boldsymbol{c}_0,\boldsymbol{c}_1) \in R^2_{Q_l, N}$ and Galois key $\mathtt{{GK}}$ as input, and it involves automorphism followed by key-switching using Galois keys. Automorphism is a special permutation of ciphertext coefficients. After automorphism, the ciphertext is encrypted under a rotated secret key. Therefore, a key-switching is required to obtain a chiphetext that is encrypted under the original secret key.  
\end{itemize}

In optimized HEAAN implementations, the ciphertexts and key-switching key are kept in the NTT domain  with RNS bases, which improves the performance. 
The NTT is a method used for efficient implementation of large-degree polynomial multiplication operation, which is detailed in Sec.~\ref{sec:background_ntt}. 
The textbook HEAAN subroutines above were presented using non-NTT representation for the sake of simplicity.
The steps of the optimized homomorphic $\mathtt{HEAAN.Add}$, $\mathtt{HEAAN.Mult}$, $\mathtt{HEAAN.Relin}$, $\mathtt{HEAAN.ModDown}$ and $\mathtt{HEAAN.Rescale}$ operations using NTT domain polynomials are given in Algorithm~\ref{algo:add_14},~\ref{algo:mult_14},~\ref{algo:relin_14},~\ref{algo:moddown_14} and~\ref{algo:rescale_14} respectively. The tilde symbol on top of a variable indicates that the polynomial is in the NTT domain.
From now on, we use the optimized subroutines with NTT representation.
%

\subsection{Number Theoretic Transform (NTT)}\label{sec:background_ntt}

The multiplication of very large degree polynomials is one of the major performance bottlenecks for the HE implementations. The Number Theoretic Transform or NTT enables fast polynomial multiplications by reducing the complexity of polynomial multiplication to $\mathcal{O}(n \cdot \log n)$, and it is extensively employed for the implementation of HE schemes.
The NTT is defined as the Discrete Fourier Transform over $\mathbb{Z}_q$. An $N$-point NTT operation transforms a polynomial $\boldsymbol{a}$ of degree $N-1$ degree polynomial into another $N-1$ degree polynomial $\boldsymbol{\tilde{a}}$ .
The NTT uses the powers of $N$-th root of unity $\omega$ (also referred to as twiddle factors) which satisfies $\omega^n \equiv 1 \pmod q$ and $\omega^i \neq 1 \pmod q$ $\forall i < N$, where $q \equiv 1 \pmod N$. 
Similarly, inverse NTT (INTT) follows the same method with the modular inverse of $\omega$ and the resulting coefficients should be scaled by $1/N$.
%

When the polynomial ring is in the form of $\mathbb{Z}_q[x]/x^N+1$ and $q \equiv 1 \pmod{2N}$, \textit{negative wrapped convolution} technique enables an efficient NTT-based polynomial multiplication method. First, NTT is performed on input polynomials, then the resulting polynomials are coefficient-wise multiplied and finally, the INTT operation is performed.
However, this technique requires input and output polynomials to be pre- and post-processed with $2N$-th root of unity, respectively.
Roy~\textit{et al.}~\cite{roy14} and Poppelmann~\textit{et al.}~\cite{poppelmann15} showed how to merge pre-processing with NTT and post-processing with INTT, respectively. These new NTT and INTT operations use $2N$-th root of unity and its modular inverse, respectively. Polynomial multiplication operation $\boldsymbol{c} = \boldsymbol{a} \cdot \boldsymbol{b} \pmod{x^N+1}$ can be performed using new NTT and INTT operations as shown in Eqn.~\ref{eqn:ntt_2}.

\begin{equation}\label{eqn:ntt_2}
    \boldsymbol{c} = \mathtt{INTT}(\mathtt{NTT}(\boldsymbol{a}) \star \mathtt{NTT}(\boldsymbol{b}))
\end{equation}

%

%% file: 3.flex_technique.tex
\section{Flexible Design Methodology} \label{sec:flex}
\subsection{Homomorphic Encryption Applications Require Flexibility}
Different homomorphic applications require different multiplicative depths, hence the parameter sets of the underlying homomorphic encryption instances change with the complexity of the application. 
For example, the SEAL~\cite{SEAL3.6} software library uses RNS-HEAAN parameters $(\log_2Q,N)= (109, 2^{12})$, $(218, 2^{13})$ and $(438, 2^{14})$ for the multiplicative depths 2, 4, and 8, respectively. 
%
As software implementations are inherently flexible, such changes in the parameter sets do not require a re-implementation of the software -- they affect only the performance and memory allocation overheads. E.g., SEAL~\cite{SEAL3.6} handles the memory allocation for different parameters flexibly by dynamically allocating memory blocks in the heap. On the other hand, hardware implementations are generally not flexible as the compute cores and memory elements are \emph{hardwired} by a fixed set of wires. As a consequence, when the hardware acceleration of cloud-side homomorphic evaluation routines is considered, the drastic changes in the sizes of the parameter sets pose a major challenge. To satisfy the flexibility requirements associated with different applications, one naive approach will be to install several parameter-specific optimized accelerators in the cloud. With that, the cloud can choose an appropriate accelerator for evaluating a given homomorphic application. This approach will give optimal execution time for any application, but will cause redundancy by requiring several accelerator units. A second approach will be to use a single accelerator which has been designed for a large parameter. With this approach, all homomorphic applications irrespective of their complexities will have to use the large and fixed parameter set. Therefore, this approach will cause poor performances for applications that do not require a large parameter set. In this paper, we propose a third approach which is a divide-and-conquer approach. It enables using a single hardware accelerator, that has been designed for a smaller parameter, for supporting several larger parameter sets flexibly without causing any major performance loss.

For the cloud-side operations, the size of coefficient modulus $Q$ and the polynomial degree $N$ change with the multiplicative depth. Attaining flexibility in the modulus size is relatively easy if the homomorphic encryption scheme uses RNS.   
As explained in Sec.~\ref{sec:background_heaan}, an RNS-based homomorphic encryption scheme operates using a set of moduli. %
Therefore, a change in the size of the coefficient modulus is equivalent to addition or deletion of moduli to/from the RNS base. 
The hardware accelerators~\cite{roy_hpca, he_aws} implement $Q$-flexibility by instantiating several parallel residue arithmetic processors in the architecture. Each such processor is used in a time-shared manner to support more than one moduli.

Implementing flexibility for the polynomial degree $N$ in a hardware accelerator is not straightforward.  
In the literature, there is only one accelerator F1~\cite{feldmann_2021f1} that supports multiple parameter sets with the polynomial degrees ranging from $2^{10}$ to $2^{14}$ by allocating  computational resources for the largest parameter. For smaller parameters, it uses a subset of the available computational resources.
This approach increases the complexity of the implementation as it requires the computational resources for the largest parameter. Also, extra logic elements are required to bypass a subset of idle computational resources for smaller parameters. %

Our proposed flexible accelerator design methodology is fundamentally different from the method used in F1. It uses arithmetic units that have been optimized for a fixed-and-smaller parameter to perform operations for larger parameters. %
\color{black}

\subsection{Design Methodology for Flexibility in Polynomial Degree}\label{sec:flex_method}

Let's assume that there is an accelerator architecture that has been optimized for the homomorphic operations with polynomial degree $N$. The research problem is: How could we use the accelerator for larger parameter sets, e.g., with the polynomial degrees $2N$, $4N$, etc.?  
To solve the problem, we propose a divide-and-conquer approach. We apply the polynomial version of CRT to the ciphertexts (which consist of residue polynomials) and split a ciphertext with a large polynomial degree (i.e., $ \mathtt{ct} \in R^2_{Q,2N}$) into several ciphertexts with smaller polynomial degrees (i.e., $ \mathtt{ct'} \in R^4_{Q,N}$).
%
Next, these smaller ciphertexts (with polynomial degrees $N$) 
are processed using the given hardware accelerator that has been optimized for the polynomial degree $N$.
The splitting can be applied recursively.

Let the irreducible polynomial $\boldsymbol{f}$  of the ring $R_{Q,2N} = \mathbb{Z}_{Q}[x]/\boldsymbol{f}$ be of the form $\boldsymbol{f}= x^{2N} + 1$.  Let all the moduli $q_i$ in the RNS base of the $Q$ be primes such that $q_i \equiv 1 \pmod{4N}$. Then $4N$-th primitive root of the unity will exist in $\mathbb{Z}_{q_i}$. Let $\zeta_{4N}$ be one such primitive root of the unity, i.e., $\zeta_{4N}^{4N} \equiv 1 \pmod{q_i}$ and $\zeta_{4N}^{2N} \equiv -1 \pmod{q_i}$.
Then, $\boldsymbol{f}$ can be re-written as $x^{2N}-\zeta_{4N}^{2N}$ and factored into polynomials $\boldsymbol{f}_p$ and $\boldsymbol{f}_m$ in $\mathbb{Z}_{Q}$ as shown in Eqn.~\ref{eqn:sj3}.
\begin{equation}\label{eqn:sj3}
    \boldsymbol{f}(x) = \boldsymbol{f}_p(x) \cdot \boldsymbol{f}_m(x) = (x^{N}-\zeta_{4N}^{N}) \cdot (x^{N}+\zeta_{4N}^{N})
\end{equation}

Note that $\boldsymbol{f}$ is irreducible over integers but not over $\mathbb{Z}_{Q}$.
Since $\boldsymbol{f}_p$ and $\boldsymbol{f}_m$ are coprime, the isomorphism $R_{q_i,2N} = \mathbb{Z}_{q_i}[x]/\boldsymbol{f} \mapsto \mathbb{Z}_{q_i}[x]/\boldsymbol{f}_p \times \mathbb{Z}_{q_i}[x]/\boldsymbol{f}_m$ exists.
Thus, any operation in the polynomial ring $R_{q_i,2N}$ can be mapped to smaller operations in the polynomial rings $\mathbb{Z}_{q_i}[x]/\boldsymbol{f}_p$ and $\mathbb{Z}_{q_i}[x]/\boldsymbol{f}_m$.
From now on, we use $R_{q_i,N,p}$ and $R_{q_i,N,m}$ (\textit{cf.} $R_{Q,N,p}$ and $R_{Q,N,m}$) to represent the polynomial rings $\mathbb{Z}_{q_i}[x]/\boldsymbol{f}_p$ and $\mathbb{Z}_{q_i}[x]/\boldsymbol{f}_m$ (\textit{cf.} $\mathbb{Z}_{Q}[x]/\boldsymbol{f}_p$ and $\mathbb{Z}_{Q}[x]/\boldsymbol{f}_m$), respectively.

\noindent\begin{minipage}[t]{0.49\textwidth}
    \begin{algorithm}[H]
    \renewcommand{\algorithmicrequire}{\textbf{In:}}
    \renewcommand{\algorithmicensure}{\textbf{Out:}}
    \caption{$\mathtt{HEAAN.Add}$ Algorithm for $2N$}
    \begin{algorithmic}[1]

        \Require $\mathtt{ct}^{(p)}=(\tilde{\boldsymbol{c}}_{0}^{(p)}, \tilde{\boldsymbol{c}}_{1}^{(p)}) \in R_{Q_l,N,p}^2$ 
        \Require $\mathtt{ct}^{(m)}=(\tilde{\boldsymbol{c}}_{0}^{(m)}, \tilde{\boldsymbol{c}}_{1}^{(m)}) \in R_{Q_l,N,m}^2$ 
        
        \Require $\mathtt{ct}'^{(p)}=(\tilde{\boldsymbol{c}}_{0}'^{(p)}, \tilde{\boldsymbol{c}}_{1}'^{(p)}) \in R_{Q_l,N,p}^2$ 
        \Require $\mathtt{ct}'^{(m)}=(\tilde{\boldsymbol{c}}_{0}'^{(m)}, \tilde{\boldsymbol{c}}_{1}'^{(m)}) \in R_{Q_l,N,m}^2$

        \Ensure  $\mathtt{d}^{(p)}=(\tilde{\boldsymbol{d}}_{0}^{(p)}, \tilde{\boldsymbol{d}}_{1}^{(p)}) \in R_{Q_l,N,p}^{2}$
        \Ensure  $\mathtt{d}^{(m)}=(\tilde{\boldsymbol{d}}_{0}^{(m)}, \tilde{\boldsymbol{d}}_{1}^{(m)}) \in R_{Q_l,N,m}^{2}$

    	\State $\tilde{\boldsymbol{d}}_{0}^{(p)} \leftarrow \tilde{\boldsymbol{c}}_{0}^{(p)} + \tilde{\boldsymbol{c}}_{0}'^{(p)}$ 
    	\State $\tilde{\boldsymbol{d}}_{1}^{(p)} \leftarrow \tilde{\boldsymbol{c}}_{1}^{(p)} + \tilde{\boldsymbol{c}}_{1}'^{(p)}$
    	
    	\State $\tilde{\boldsymbol{d}}_{0}^{(m)} \leftarrow \tilde{\boldsymbol{c}}_{0}^{(m)} + \tilde{\boldsymbol{c}}_{0}'^{(m)}$ 
    	\State $\tilde{\boldsymbol{d}}_{1}^{(m)} \leftarrow \tilde{\boldsymbol{c}}_{1}^{(m)} + \tilde{\boldsymbol{c}}_{1}'^{(m)}$
    \end{algorithmic}\label{algo:homadd_15}
    \end{algorithm}
\end{minipage}
\hfill
\begin{minipage}[t]{0.49\textwidth}
    \begin{algorithm}[H]
    \renewcommand{\algorithmicrequire}{\textbf{In:}}
    \renewcommand{\algorithmicensure}{\textbf{Out:}}
    \caption{$\mathtt{HEAAN.Mult}$ Algorithm for $2N$}
    \begin{algorithmic}[1]
        \Require $\mathtt{ct}^{(p)}=(\tilde{\boldsymbol{c}}_{0}^{(p)}, \tilde{\boldsymbol{c}}_{1}^{(p)}) \in R_{Q_l,N,p}^2$ 
        \Require $\mathtt{ct}^{(m)}=(\tilde{\boldsymbol{c}}_{0}^{(m)}, \tilde{\boldsymbol{c}}_{1}^{(m)}) \in R_{Q_l,N,m}^2$ 
        
        \Require $\mathtt{ct}'^{(p)}=(\tilde{\boldsymbol{c}}_{0}'^{(p)}, \tilde{\boldsymbol{c}}_{1}'^{(p)}) \in R_{Q_l,N,p}^2$ 
        \Require $\mathtt{ct}'^{(m)}=(\tilde{\boldsymbol{c}}_{0}'^{(m)}, \tilde{\boldsymbol{c}}_{1}'^{(m)}) \in R_{Q_l,N,m}^2$

        \Ensure  $\mathtt{d}^{(p)}=(\tilde{\boldsymbol{d}}_{0}^{(p)}, \tilde{\boldsymbol{d}}_{1}^{(p)}, \tilde{\boldsymbol{d}}_{2}^{(p)}) \in R_{Q_l,N,p}^{3}$
        \Ensure  $\mathtt{d}^{(m)}=(\tilde{\boldsymbol{d}}_{0}^{(m)}, \tilde{\boldsymbol{d}}_{1}^{(m)}, \tilde{\boldsymbol{d}}_{2}^{(m)}) \in R_{Q_l,N,m}^{3}$

        \State $\tilde{\boldsymbol{d}}_{0}^{(p)} \leftarrow \tilde{\boldsymbol{c}}_{0}^{(p)} \star \tilde{\boldsymbol{c}}_{0}'^{(p)}$, $\tilde{\boldsymbol{d}}_{2}^{(p)} \leftarrow \tilde{\boldsymbol{c}}_{1}^{(p)} \star \tilde{\boldsymbol{c}}_{1}'^{(p)}$ 
        \State $\tilde{\boldsymbol{d}}_{1}^{(p)} \leftarrow \tilde{\boldsymbol{c}}_{0}^{(p)} \star \tilde{\boldsymbol{c}}_{1}'^{(p)} + \tilde{\boldsymbol{c}}_{1}^{(p)} \star \tilde{\boldsymbol{c}}_{0}'^{(p)}$
        
        \State $\tilde{\boldsymbol{d}}_{0}^{(m)} \leftarrow \tilde{\boldsymbol{c}}_{0}^{(m)} \star \tilde{\boldsymbol{c}}_{0}'^{(m)}$, $\tilde{\boldsymbol{d}}_{2}^{(m)} \leftarrow \tilde{\boldsymbol{c}}_{1}^{(m)} \star \tilde{\boldsymbol{c}}_{1}'^{(m)}$ 
        \State $\tilde{\boldsymbol{d}}_{1}^{(m)} \leftarrow \tilde{\boldsymbol{c}}_{0}^{(m)} \star \tilde{\boldsymbol{c}}_{1}'^{(m)} + \tilde{\boldsymbol{c}}_{1}^{(m)} \star \tilde{\boldsymbol{c}}_{0}'^{(m)}$
    \end{algorithmic}\label{algo:hommul_15}
    \end{algorithm}
\end{minipage}
\vspace{1em}

Using the isomorphism a large ciphertext $\mathtt{ct} = (\boldsymbol{c}_0, \boldsymbol{c}_1) \in R_{Q,2N}^2$ can be split into two smaller ciphertexts $\mathtt{ct}^{(p)} = (\boldsymbol{c}_0^{(p)}, \boldsymbol{c}_1^{(p)}) \in R_{Q,N,p}^2$ and $\mathtt{ct}^{(m)} = (\boldsymbol{c}_0^{(m)}, \boldsymbol{c}_1^{(m)}) \in R_{Q,N,m}^2$. 
After that, the homomorphic operations can be performed on the smaller ciphertexts.
This approach enables efficient re-use of a fixed hardware architecture that has been optimized for the smaller parameter with the polynomial degree $N$. Additionally, management of the smaller data variables in the limited on-chip memory of a hardware accelerator becomes relatively easier. %

The splitting of a ciphertext is essentially the splittings of its polynomials. A polynomial $\boldsymbol{a} \in R_{Q,2N}$ is split into the two polynomials $\boldsymbol{a}_p = \boldsymbol{a} \pmod{\boldsymbol{f}_p} \in R_{Q,N,p}$ and $\boldsymbol{a}_m = \boldsymbol{a} \pmod{\boldsymbol{f}_m} \in R_{Q,N,m}$. On the opposite mapping, the two smaller polynomials $\boldsymbol{a}_p$ and $\boldsymbol{a}_m$ can be joined using the CRT to obtain the large polynomial $\boldsymbol{a}$ as 
\begin{equation}
\boldsymbol{a} = (\boldsymbol{a}_p+\boldsymbol{a}_m)\frac{1}{2} + 
(\boldsymbol{a}_p-\boldsymbol{a}_m)\frac{\zeta_{4N}^{-N}}{2} x^N \pmod{Q}.    
\end{equation}

The same applies to the key-switching key. The splitting and joining operations each have a linear time complexity. They can be performed in the software or in the hardware. For example, a software system that is using a hardware accelerator as a coprocessor, can do the splitting job before sending the small ciphertexts to the hardware. Similarly, the software system can combine the partial results returned by the hardware. 
In the remaining part of the paper, we will use the two functions $\mathtt{Split()}$ and $\mathtt{Join()}$ to perform the splitting and joining operations respectively.

\subsection{RNS-HEAAN Subroutines with Flexible Method}\label{sec:flex_routines}

Here we provide concrete algorithmic descriptions of the subroutines of RNS-HEAAN to implement the flexibility that we discussed in the previous section. \newline

\noindent\textbf{$\mathtt{HEAAN.Add}$ and $\mathtt{HEAAN.Mult}$ subroutines:} 
Let two large ciphertexts be $\mathtt{ct}$ and $\mathtt{ct'} \in R_{Q_l,2N}^2$. Using the $\mathtt{Split}()$ function, each of them is split into two smaller ciphertexts. 
Algorithm~\ref{algo:homadd_15} and Algorithm~\ref{algo:hommul_15} show how the smaller ciphertexts are processed to perform the homomorphic addition and multiplication in the larger ring $R_{Q_l,2N}$. In the RNS these polynomial operations are performed on the residue polynomials. 
\newline

\begin{algorithm}[t!]
\renewcommand{\algorithmicrequire}{\textbf{In:}}
\renewcommand{\algorithmicensure}{\textbf{Out:}}
\caption{$\mathtt{HEAAN.Relin}$ Algorithm for $2N$}
\begin{algorithmic}[1]
    \Require  $\mathtt{d}^{(p)} = (\tilde{\boldsymbol{d}}_{0}^{(p)}, \tilde{\boldsymbol{d}}_{1}^{(p)}, \tilde{\boldsymbol{d}}_{2}^{(p)}) \in R_{Q_l,N,p}^{3}$, $\mathtt{d}^{(m)} = (\tilde{\boldsymbol{d}}_{0}^{(m)}, \tilde{\boldsymbol{d}}_{1}^{(m)}, \tilde{\boldsymbol{d}}_{2}^{(m)}) \in R_{Q_l,N,m}^{3}$ 
        
    \Require  $\tilde{\mathtt{KSK}}_{0}^{(p)} \in R_{pQ_l,N,p}^l$, $\tilde{\mathtt{KSK}}_{1}^{(p)} \in R_{pQ_l,N,p}^l$, $\tilde{\mathtt{KSK}}_{0}^{(m)} \in R_{pQ_l,N,m}^l$, $\tilde{\mathtt{KSK}}_{1}^{(m)} \in R_{pQ_l,N,m}^l$
    
    \Ensure   $\mathtt{d}^{(p)} = (\tilde{\boldsymbol{d}}_{0}^{(p)}, \tilde{\boldsymbol{d}}_{1}^{(p)}) \in R_{Q_l,N,p}^{2}$, $\mathtt{d}^{(m)} = (\tilde{\boldsymbol{d}}_{0}^{(m)}, \tilde{\boldsymbol{d}}_{1}^{(m)}) \in R_{Q_l,N,m}^{2}$
    
    \For{$j=0$ to $l-1$}	
	
	\State $\boldsymbol{d}_{2}^{(p)}[j] \leftarrow \mathtt{INTT_p}(\tilde{\boldsymbol{d}}_{2}^{(p)}[j])$, $\boldsymbol{d}_{2}^{(m)}[j] \leftarrow \mathtt{INTT_m}(\tilde{\boldsymbol{d}}_{2}^{(m)}[j])$ \Comment{in $\mathbb{Z}_{q_j}$}
    \State $\boldsymbol{d}_{2}[j] \leftarrow \mathtt{Join}(\boldsymbol{d}_{2}^{(p)}[j], \boldsymbol{d}_{2}^{(m)}[j])$

    \EndFor

    \For{$j=0$ to $l$}	\Comment{Here $q_{l}$ is used to represent $p$}	
	    \State $(\boldsymbol{c}_{0}''^{(p)}[j], \boldsymbol{c}_{1}''^{(p)}[j], \boldsymbol{c}_{0}''^{(m)}[j], \boldsymbol{c}_{1}''^{(m)}[j]) \leftarrow 0$

	\For{$i=0$ to $l-1$}	
    	\State $\boldsymbol{r}^{(p)}, \boldsymbol{r}^{(m)} \leftarrow \mathtt{Split}(\big[\boldsymbol{d}_{2}[i]\big]_{q_j})$
    	
    	\State $\tilde{\boldsymbol{r}}^{(p)} \leftarrow \mathtt{NTT_p}(\boldsymbol{r}^{(p)})$, $\tilde{\boldsymbol{r}}^{(m)} \leftarrow \mathtt{NTT_m}(\boldsymbol{r}^{(m)})$ \Comment{in $\mathbb{Z}_{q_j}$} 
  	
    	\State $\boldsymbol{c}_{0}''^{(p)}[j] \leftarrow \big[\boldsymbol{c}_{0}''^{(p)}[j] + \tilde{\mathtt{KSK}}_{0}^{(p)}[i][j] \star \tilde{\boldsymbol{r}}^{(p)}\big]_{q_j}$
    	\State $\boldsymbol{c}_{1}''^{(p)}[j] \leftarrow \big[\boldsymbol{c}_{1}''^{(p)}[j] + \tilde{\mathtt{KSK}}_{1}^{(p)}[i][j] \star \tilde{\boldsymbol{r}}^{(p)}\big]_{q_j}$
    	
    	\State $\boldsymbol{c}_{0}''^{(m)}[j] \leftarrow \big[\boldsymbol{c}_{0}''^{(m)}[j] + \tilde{\mathtt{KSK}}_{0}^{(m)}[i][j] \star \tilde{\boldsymbol{r}}^{(m)}\big]_{q_j}$ 
    	\State $\boldsymbol{c}_{1}''^{(m)}[j] \leftarrow \big[\boldsymbol{c}_{1}''^{(m)}[j] + \tilde{\mathtt{KSK}}_{1}^{(m)}[i][j] \star \tilde{\boldsymbol{r}}^{(m)}\big]_{q_j}$
    	
  	
	\EndFor	
	
	\EndFor

    \State $(\tilde{\boldsymbol{d}}_{0}'^{(p)}, \tilde{\boldsymbol{d}}_{0}'^{(m)}) \leftarrow (\tilde{\boldsymbol{d}}_{0}^{(p)}, \tilde{\boldsymbol{d}}_{0}^{(m)}) + (\mathtt{HEAAN.ModDown}(\tilde{\boldsymbol{c}}_{0}''^{(p)}), \mathtt{HEAAN.ModDown}(\tilde{\boldsymbol{c}}_{0}''^{(m)}))$
    
    \State $(\tilde{\boldsymbol{d}}_{1}'^{(p)}, \tilde{\boldsymbol{d}}_{1}'^{(m)}) \leftarrow (\tilde{\boldsymbol{d}}_{1}^{(p)}, \tilde{\boldsymbol{d}}_{1}^{(m)}) + (\mathtt{HEAAN.ModDown}(\tilde{\boldsymbol{c}}_{1}''^{(p)}), \mathtt{HEAAN.ModDown}(\tilde{\boldsymbol{c}}_{1}''^{(m)}))$
   	
\end{algorithmic}\label{algo:relin_15}
\end{algorithm} 

\begin{algorithm}[t!]
\renewcommand{\algorithmicrequire}{\textbf{In:}}
\renewcommand{\algorithmicensure}{\textbf{Out:}}
\caption{$\mathtt{HEAAN.Rescale}$ Algorithm for $2N$}
\begin{algorithmic}[1]
    \Require  $\tilde{\boldsymbol{d}}^{(p)} \in R_{Q_l,N,p}, \tilde{\boldsymbol{d}}^{(m)} \in R_{Q_l,N,m}$
    
    \Ensure $\tilde{\boldsymbol{d}}'^{(p)} \in R_{Q_{l-1},N,p}, \tilde{\boldsymbol{d}}'^{(m)} \in R_{Q_{l-1},N,m}$
    
    \State $\boldsymbol{t}^{(p)} \leftarrow \mathtt{INTT_p}(\tilde{\boldsymbol{d}}^{(p)}[l-1])$, $\boldsymbol{t}^{(m)} \leftarrow \mathtt{INTT_m}(\tilde{\boldsymbol{d}}^{(m)}[l-1])$
    
    \State $\boldsymbol{t} \leftarrow \mathtt{Join}(\boldsymbol{t}^{(p)}, \boldsymbol{t}^{(m)})$
    
    \For{$i=0$ to $l-2$}	
    
    \State $\boldsymbol{r}^{(p)}, \boldsymbol{r}^{(m)} \leftarrow \mathtt{Split}(\big[\boldsymbol{t}\big]_{q_i})$
	
	\State $\tilde{\boldsymbol{r}}^{(p)} \leftarrow \mathtt{NTT_p}(\boldsymbol{r}^{(p)})$, $\tilde{\boldsymbol{r}}^{(m)} \leftarrow \mathtt{NTT_m}(\boldsymbol{r}^{(m)})$ \Comment{in $\mathbb{Z}_{q_i}$}
	
	\State $\tilde{\boldsymbol{d}}'^{(p)}[i] \leftarrow \big[ q^{-1}_l \odot (\tilde{\boldsymbol{d}}^{(p)}[i] - \tilde{\boldsymbol{r}}^{(p)})\big]_{q_i}$, $\tilde{\boldsymbol{d}}'^{(m)}[i] \leftarrow \big[q^{-1}_l \odot (\tilde{\boldsymbol{d}}^{(m)}[i] - \tilde{\boldsymbol{r}}^{(m)})\big]_{q_i}$

    \EndFor

\end{algorithmic}\label{algo:rescale_15}
\end{algorithm}

\noindent\textbf{$\mathtt{HEAAN.Relin}$ and $\mathtt{HEAAN.Rescale}$ subroutines:}
The steps of relinearization and rescaling operations for the polynomial degree of $2N$ are shown in Algorithm~\ref{algo:relin_15} and Algorithm~\ref{algo:rescale_15}, respectively. The mod-down and rescaling operations are very similar as shown in Algorithm~\ref{algo:moddown_14} and Algorithm~\ref{algo:rescale_14}, thus the steps of the mod-down operation for $2N$ can easily be constructed from the steps of rescaling operation (Algorithm~\ref{algo:rescale_15}).
The relinearization, mod-down and rescaling operations for the polynomial degree of $2N$ are more complex than the homomorphic addition and multiplication operations as they require the $\mathtt{Split/Join}$ and $\mathtt{NTT/INTT}$ operations. 
As shown in step 7 of Algorithm~\ref{algo:relin_14}, the key switching multiplication part of the relinearization operation requires the polynomials of the ciphertext $\boldsymbol{d}_{2}$ to be reduced by $q_j$, which cannot be performed independently on $\boldsymbol{d}_{2}^{(p)}$ and $\boldsymbol{d}_{2}^{(m)}$. Therefore, before the modular reduction by $q_j$ the two small polynomials are joined using $\mathtt{Join}$ in step 3 of Algorithm~\ref{algo:relin_15}. 
After the modular reduction, the reduced large polynomial is again split into two small polynomials using $\mathtt{Split}$ in step 8 of Algorithm~\ref{algo:relin_15}.  
Similarly, the mod-down and rescaling operations also require the split and join operations as they involve modular reductions.

Note that the resultant polynomials after a $\mathtt{Split}$ operation reside in two different polynomial rings $\mathbb{Z}_Q[x]/x^N - \zeta_{4N}^{N}$ and $\mathbb{Z}_Q[x]/x^N + \zeta_{4N}^{N}$. 
The NTT and INTT operations over these different polynomial rings are performed using different twiddle constants. We use $\mathtt{NTT_p}$/$\mathtt{INTT_p}$ and $\mathtt{NTT_m}$/$\mathtt{INTT_m}$ to represent the NTT/INTT operations over the two rings $\mathbb{Z}_Q[x]/x^N - \zeta_{4N}^{N}$ and $\mathbb{Z}_Q[x]/x^N + \zeta_{4N}^{N}$, respectively.\\

%% file: 4.low_arith_units.tex
\section{Architecture of Medha}\label{sec:medha}

\subsection{Parameter Selection}\label{sec:params}

To implement RNS-HEAAN with the proposed flexible design method, we use 60 and 54-bit prime moduli as RNS bases similar to SEAL~\cite{SEAL3.6} and HEAX~\cite{heax} with each prime satisfying $q_i \equiv 1 \pmod{4N}$. 
For proof of the concept, Medha is implemented to support two large parameter sets as shown in Table~\ref{tab:results_paramset}. 
It supports two polynomial degrees, $N=2^{14}$ and $2N=2^{15}$, with up to 546-bit coefficient modulus size using 10 prime moduli. 
Next, we present the low-level arithmetic units of the Medha.

\begin{table}[t!]
\caption{Parameter Sets}\label{tab:results_paramset}
\centering
\begin{tabular}{c | c c c c}
\hline

\textbf{Param. Set}   & $(\log_2(pQ),N)$ & $L+1$ & \textbf{Mul. Depth} & \textbf{Sec. Level}  \\ \hline\hline

Set-1 & ($438,2^{14}$) & 8  & 7  & >128-bit \\ 
Set-2 & ($546,2^{15}$) & 10 & 9  & >128-bit \\  \hline
\end{tabular}
\end{table}

\subsection{Low-level Arithmetic Units} \label{sec:imp_low_level}

In this section, we present the low-level arithmetic units of Medha starting from the lowest level of the implementation hierarchy (Fig.~\ref{fig:he_hieararchy}) containing modular arithmetic, namely addition, subtraction and multiplication. They are the most frequently used modules. We use bit-parallel multipliers made of DSPs available in FPGAs. To bypass the expensive modular reduction circuits from~\cite{roy_hpca, heax}, we use pseudo-Mersenne primes in the RNS base and perform very cheap modular reductions. Similar reduction circuits are popular in lattice-based post-quantum cryptography. 
See Appendix~\ref{app:modred} for details.

\subsubsection{Parallel NTT Architecture}\label{sec:ntt_memory_access}

The next level of Fig.~\ref{fig:he_hieararchy}, performs the arithmetic of large-degree polynomials. Polynomial multiplication is the most expensive operation and it can be implemented efficiently using NTT-based polynomial multiplication. The NTT unit has been designed and optimized for $N=2^{14}$. 

Our NTT-based multiplier uses an iterative NTT approach. It utilizes the decimation-in-time (DiT) approach for the forward NTT and decimation-in-frequency (DiF) approach for the inverse NTT (INTT)~\cite{ntt_intt}. This particular combination eliminates the need for any coefficient permutations before the NTTs or after the INTT. The DiT NTT Algorithm is presented in Algorithm~\ref{algo:dit_ntt} in Appendix~\ref{app:ntt}.

\begin{figure}[t]
\centering
\includegraphics[width=5.5in]{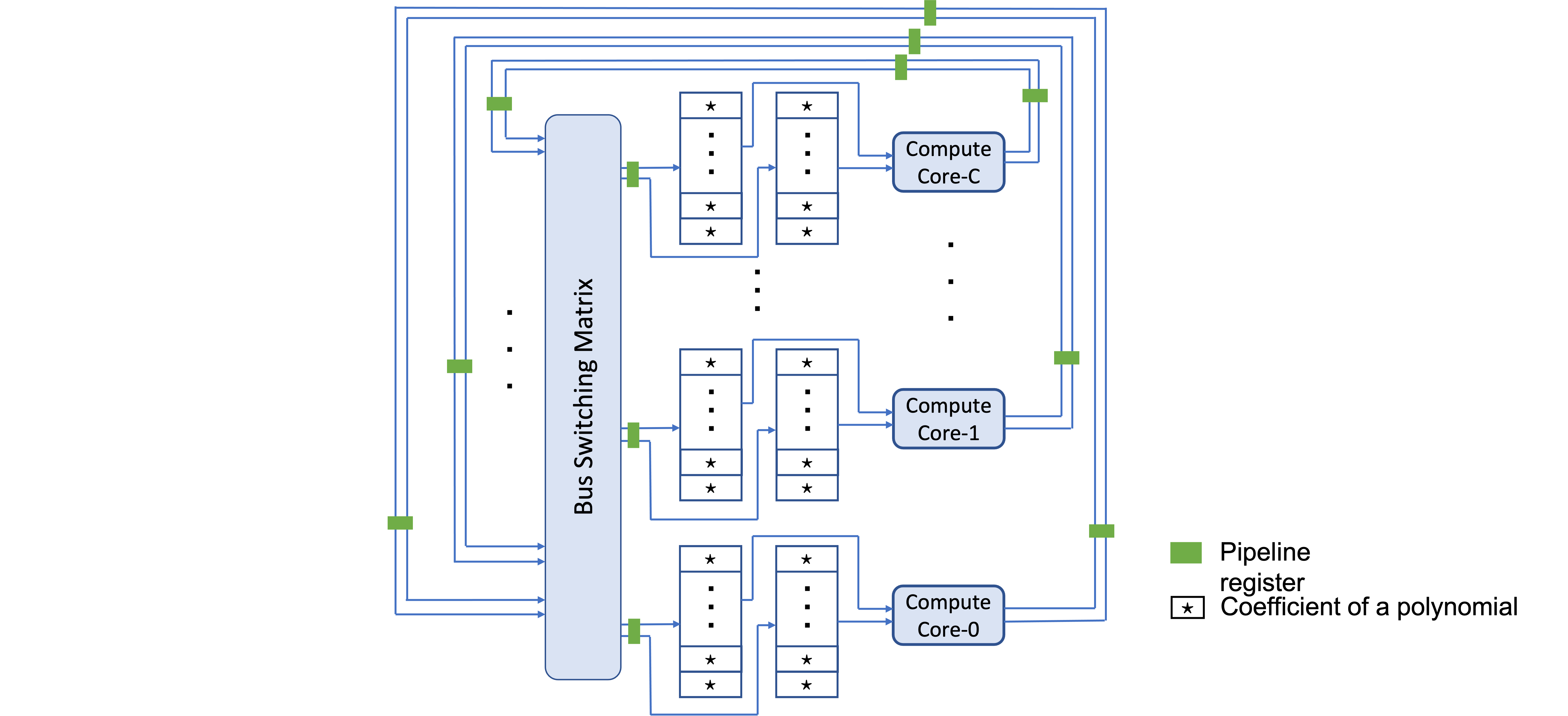}
\caption{Organization of memory and compute cores for parallel NTT}
\label{fig:memory_ntt}
\end{figure}

For fast polynomial multiplications, we implement a multi-core parallel NTT unit borrowing the best practices from~\cite{roy_tc2018,roy_hpca,heax} including the routing and BRAM access optimizations from~\cite{roy_tc2018}. In addition, we apply an `address delaying' technique that results in a significant reduction in the register consumption without causing any performance overhead. 
Fig.~\ref{fig:memory_ntt} shows a high-level organization of the memory elements (for storing the polynomial-parts) and the compute cores (for processing the coefficients) inside the NTT unit. 
The bus matrix rearranges the processed coefficients before writing them to the memory elements. 
We put multiple layers of pipeline registers (shown in green in Fig.~\ref{fig:memory_ntt}) to increase the clock frequency. The number of parallel compute cores is a design parameter that depends on area and performance budgets. Our NTT unit is optimized for polynomials of $N=2^{14}$ coefficients and it uses 16 compute cores. It computes one $2^{14}$-point NTT in around 7,168 cycles.

\subsubsection{Unified Compute Core for DiT NTT and DiF INTT}

We design a single NTT unit that employs unified compute cores for the DiT NTT and DiF INTT. 
In DiT, a new coefficient-pair $(u', t')$ is computed from $(u, t)$ as follows: $(u', t') \leftarrow (u + t \cdot \omega, u - t \cdot \omega)$. Whereas in DiF, a new coefficient-pair is computed
as $(u'', t'') \leftarrow (\frac{u + t}{2}, \frac{(u-t) \cdot \omega}{2})$. These operations are referred to as the butterfly operations.
Our unified compute core uses one modular multiplier, one modular adder, two modular subtractors and a few two-to-one multiplexers. The modular multiplier is heavily pipelined (around 20 stages) to achieve high clock frequency. From now on, we refer to compute core as the butterfly core.
\noindent\textbf{Address delaying optimization:}
We propose an `address delaying' optimization that causes a significant reduction in the register consumption of the pipelined butterfly cores.
Let us consider the processing of coefficient pairs during DiT NTT. 
After reading the coefficient $t$, computation of the intermediate data $\omega \cdot t$ progresses through a long chain of pipeline registers present inside the modular multiplier. 
For the correct computation of $(u + t \cdot \omega, u - t \cdot \omega)$, 
$u$ and $\omega \cdot t$ must reach the inputs of the adder and subtractor synchronously in the same cycle. In~\cite{roy_tc2018}, both $u$ and $t$ are read together from the memory and then $u$ is passed through a long chain of redundant registers just to make sure that both $u$ and $\omega \cdot t$ arrive together at the adder and subtracter. 
%

\begin{figure}[t]
\centering
\includegraphics[width=4.8in]{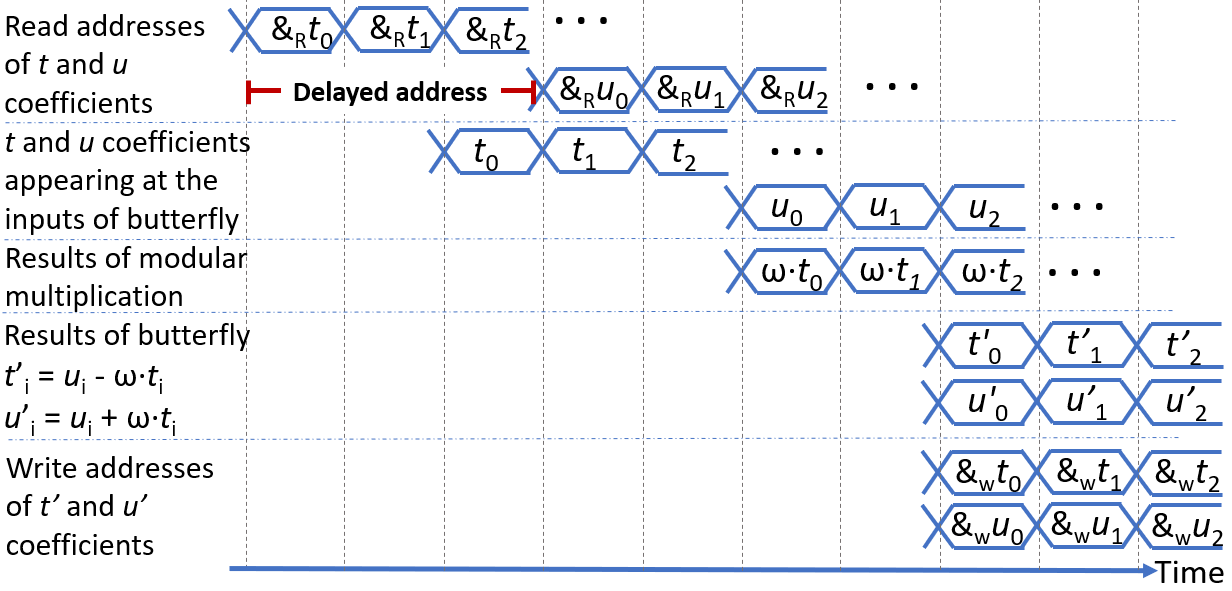}
\caption{The timing diagram for our DiT method of NTT. Due to pipelined datapath, the reading of the $u$ coefficients is delayed so that we can add or subtract them when the corresponding modular multiplication results $t \cdot \omega$ are ready. The results of butterfly operations are written synchronously. For the DiF method, the read-write happen oppositely: we read the $u$ and $t$ coefficients synchronously but write them asynchronously. The notations $\&_{R}$ and $\&_{W}$ are for reading and writing addresses respectively.}
\label{fig:NTT_timing_diagram}
\end{figure}

Our pipeline strategy avoids the above-mentioned bloated register consumption by simply delaying the read of coefficient $u$ from the memory. This technique saves around 1,200 registers per butterfly unit and around 192,00 registers for overall architecture.
%
We keep $u$ and $t$ in separate memory elements as shown in Fig.~\ref{fig:memory_ntt} so that they can be read separately just-in-time.    
The timing diagram in Fig.~\ref{fig:NTT_timing_diagram} shows how the $\{u,t\}$ coefficients are read during a DiT NTT. Reading of the $t$-coefficients for the consecutive butterfly operations is initiated several cycles (equal to the number of pipeline stages in the modular multiplier) ahead of reading the $u$-coefficients. 
As a consequence, each modular multiplication result and the respective $u$ appear synchronously at the inputs of the adder and subtractor circuits for correct computation. 
We extend the above-mentioned pipeline strategy to the DiF method of INTT. The difference is that both $u$ and $t$ are read simultaneously but the result coefficients are written separately into the memory. 

\subsubsection{Twiddle Factors during NTT}\label{subsec:twiddle_unit}
Optimized software implementations (e.g., SEAL~\cite{SEAL3.6}) and also the hardware implementations~\cite{roy_hpca, heax} save cycle count of NTT by keeping all twiddle constants in large tables.
As previous hardware implementations~\cite{roy_hpca,heax} indicated that HE is memory-bound, we follow a memory-conservative design approach and compute the twiddle constants on the fly and in parallel to the butterfly operations, therefore avoiding large BRAM consumption. %
Each NTT butterfly unit is coupled with a twiddle factor generation unit which mainly consists of a modular multiplier and two BRAM36K memory units for storing a few initial twiddle factor constants. The multiplier in the  twiddle factor generation unit is not an additional cost as it is reused to parallelize coefficient-wise multiplication and modular reduction operations.
\subsubsection{Automorphism Unit}

The automorphism operation requires permutation of polynomial coefficients.
In our architecture, the permutation is a simple memory read-write operation. Since, a HEAAN ciphertext is in the NTT domain, special care has to be taken to generate the read addresses required during the permutation. The read and write addresses are generated on the fly using a Galois element. 

\subsubsection{Building Blocks for Supporting Parameter Flexibility}

The proposed flexible design method requires very few specific low-level building blocks, which can be implemented easily. \\

\noindent\textbf{Split and Join operations:}
Although they can be performed by a software system (master mode), we implement them in the hardware to avoid unnecessary HW-SW data transfers. 
A naive approach is utilizing separate computation units for these operations.
Instead, for implementing the split and join operations efficiently, we leverage the resemblance of split and join operations to DiT and DiF butterfly structures respectively, and implement the split and join operations without any additional hardware cost using the butterfly cores. 
Since the NTT unit has 16 butterfly cores, each of $\mathtt{Split}$ and $\mathtt{Join}$ takes $\approx$1,024 clock cycles. \\

\noindent\textbf{On-the-fly twiddle factor generation:} 
As explained in Sec.~\ref{sec:flex_method}, the proposed architecture should support $N$-point NTT and INTT operations for three different polynomial rings: $\mathbb{Z}_Q[x]/x^{2^{14}}+1$ for the parameter with the polynomial degree $N=2^{14}$; $\mathbb{Z}_Q[x]/x^{2^{14}}+\zeta_{4N}^{2N}$ and $\mathbb{Z}_Q[x]/x^{2^{14}}-\zeta_{4N}^{2N}$ for the parameter with the polynomial degree $2N=2^{15}$. The ring-specific NTTs use different twiddle constants. 
As described earlier, our twiddle factor generation unit stores only a few initial constants and starting from them computes the remaining constants on-the-fly. 
As there are three rings, three sets of initial twiddle constants are required to be stored in the memory. That will increase the on-chip memory requirement. 
Instead, we store only the initial twiddle constants for the polynomial rings $\mathbb{Z}_Q[x]/x^{2^{14}}-\zeta_{4N}^{2N}$ and $\mathbb{Z}_Q[x]/x^{2^{14}}+\zeta_{4N}^{2N}$. 
When an NTT/INTT for the ring $\mathbb{Z}_Q[x]/x^{2^{14}}+1$ is to be computed, the required initial twiddle constants are generated by multiplying those for the ring $\mathbb{Z}_Q[x]/x^{2^{14}}-\zeta_{4N}^{2N}$ with $\zeta_{4N}^i$ where $i$ changes with the NTT stages.  

%% file: 5.high_arith_units.tex
\subsection{Architecture of the Homomorphic Processor} \label{sec:imp_rpau}

We take the optimized polynomial arithmetic units from the previous section and organize them in the architecture of Medha. %
Medha is an instruction-set architecture (ISA) and therefore programmable to run the homomorphic subroutines, namely addition/subtraction, multiplication, key-switching, relinearization and rescaling flexibly by reusing the same compute units. 
%
Previous works~\cite{roy_tc2018,roy_hpca} presented ISAs for accelerating homomorphic encryption, but their performance advantages remained limited, e.g., only one order speedup~\cite{roy_hpca} compared to SW implementations. 
On the contrary, HEAX~\cite{heax} organized its polynomial arithmetic elements in a subroutine-specific manner to implement a key-switching unit that gives over two orders higher throughput compared to SW. The one order speedup of HEAX over~\cite{roy_hpca} demonstrated that block-pipelined and subroutine-specific HE accelerators are much superior to programmable accelerators that reuse compute elements.  

We organize compute and memory elements pragmatically and realize a \emph{flexible} and \emph{programmable} accelerator that achieves up to $78\times$ performance improvement compared to Microsoft SEAL~\cite{SEAL3.6}.
Our work shows that HE accelerators do not have to sacrifice programmability to achieve high speed. The following subsections describe how we organize compute and memory elements.    
%

%

\begin{figure}[t]
\centering
\includegraphics[width=4.5in]{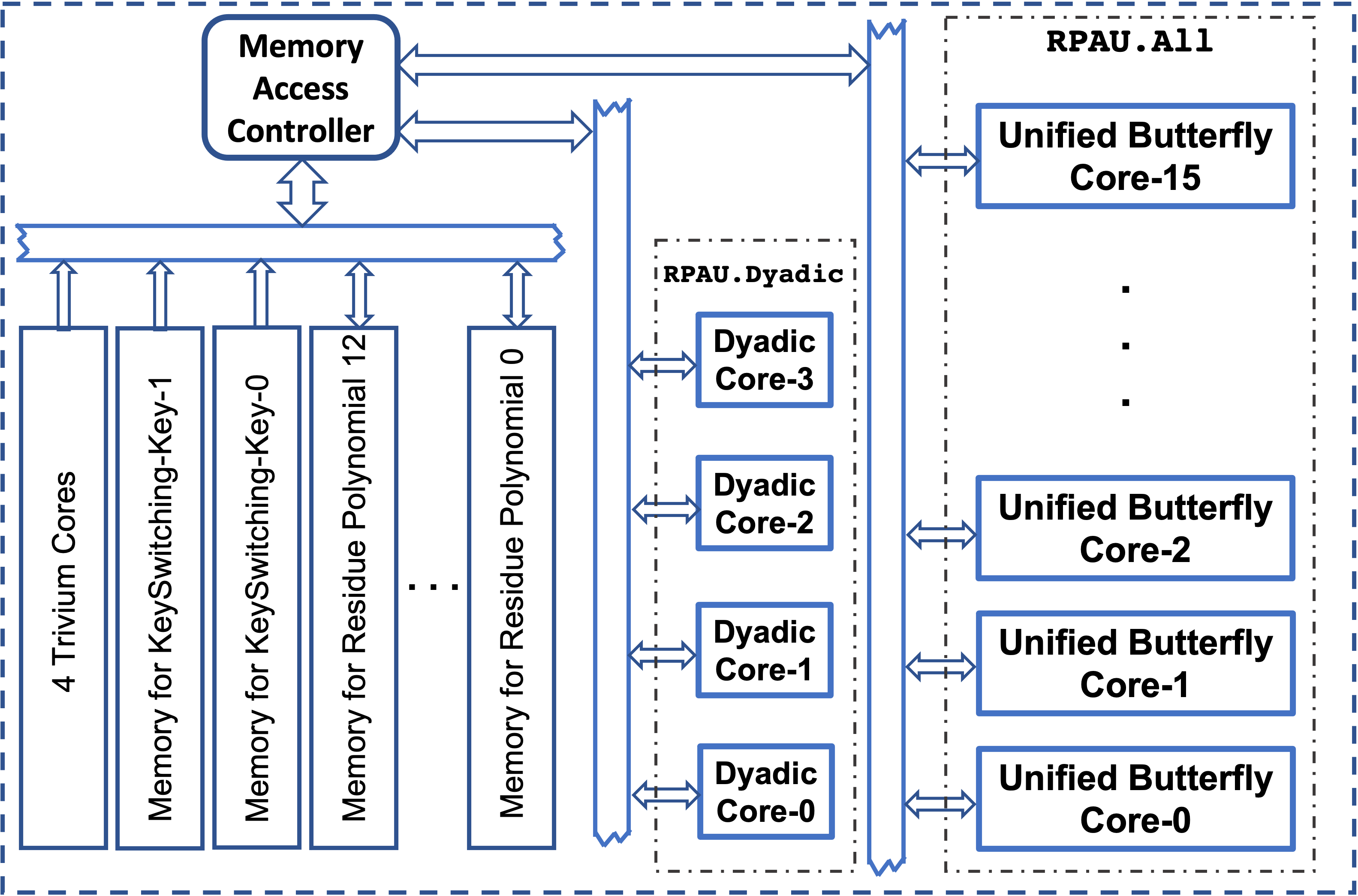}
\caption{Architecture of the Residue Polynomial Arithmetic Unit (RPAU). }
\label{fig:rpau}
\end{figure}

\subsubsection{Design of Residue Polynomial Arithmetic Unit (RPAU) }
In any RNS-based HE, an arithmetic operator is applied to a `vector' of residue polynomials. The idea has some similarities with the Single Instruction Multiple Data (SIMD) processors. To benefit from such arithmetic parallelism~\cite{roy_ches2015}, Medha instantiates multiple high-level units for processing the residue polynomials in parallel. These units are called the Residue Polynomial Arithmetic Unit (RPAU) and they are ISA. Any high-level instruction for Medha essentially gets translated into instructions for the RPAUs.

Fig.~\ref{fig:rpau} shows the organization of polynomial arithmetic cores and memory elements inside our proposed RPAU. 
%
We observe that the inner loop of key-switching or re-linearization (see Sec.~\ref{sec:background}) executes one NTT and several coefficient-wise polynomial operations. Therefore, we keep two groups of compute cores, namely \texttt{RPAU.All} and \texttt{RPAU.Dyadic} in the RPAU. 
The \texttt{RPAU.All} group is capable of performing all kinds of polynomial arithmetic operations fast using 16 cores. 
The \texttt{RPAU.Dyadic} group can only perform coefficient-wise (i.e., dyadic) operations using only 4 cores. The two compute groups are executed in parallel during key-switching (or re-linearization) and re-scaling (or mod-down) operations. \newline

\begin{figure}[t]
\centering
\includegraphics[width=5.2in]{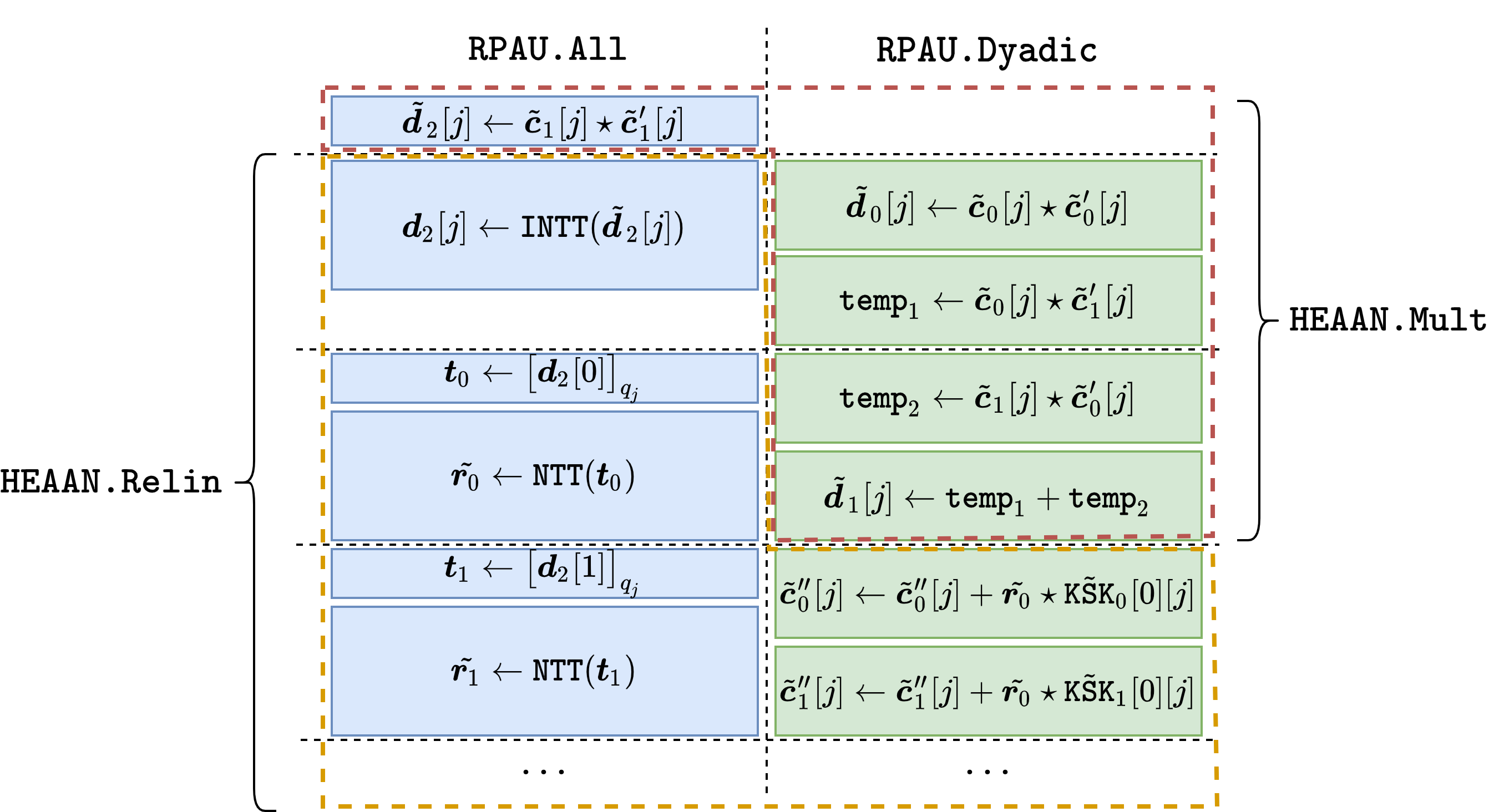}
\caption{Parallel processing of \texttt{HEAAN.Mult} and \texttt{HEAAN.Relin} using two threads inside $j$-th RPAU.}
\label{fig:mult_steps_parallel}
\end{figure}

\noindent\textbf{Parallel execution of \texttt{RPAU.All} and \texttt{RPAU.Dyadic}:}
%
Our RPAU can execute two instructions, one using \texttt{RPAU.All} and the other using \texttt{RPAU.Dyadic}, concurrently in parallel and save around 40\% cycles. The rationale behind this novel design decision is explained as follows. 
%
%

As shown in Algorithm~\ref{algo:mult_14} and Algorithm~\ref{algo:relin_14}, some of the steps in \texttt{HEAAN.Mult} and \texttt{HEAAN.Relin} can be performed in parallel. For example, calculation of $\tilde{\boldsymbol{d}}_0$, $\tilde{\boldsymbol{d}}_1$ and $\tilde{\boldsymbol{d}}_2$, $\mathtt{NTT}(\tilde{\boldsymbol{d}}_2)$ can performed using \texttt{RPAU.Dyadic} and \texttt{RPAU.All}, respectively, in parallel. 
Inside the loop of the key-switching, the steps are sequential due to data dependencies. As the loop iterates several times, we can unroll it and then `block-pipeline' the loop-internals. We have different options for applying parallel processing. %

\begin{itemize}
    \item Option 1: Running more than one NTTs in parallel inside the RPAU will be useful if we unroll the key-switching loop shown in steps 4-10 of Algorithm~\ref{algo:relin_14}.
    E.g., unrolling by a factor 4 will require 4 NTT units, therefore almost increasing the logic count of RPAU by 4 times. Hence, this approach is not attractive.
    %
    \item Option 2: Using only one NTT unit with more compute cores. Compared to the previous option, this option will be simpler as well as more effective in reducing the latency irrespective of data dependencies.  %
    E.g., instead of using 16 cores in the NTT, if we use 32 or 64 cores then we can reduce the cycle count of an NTT by a factor of 2 or 4 respectively. A potential problem is that we may not see a similar reduction in the overall computation \emph{time} due to a slow-down in the clock frequency of the much larger architecture. Another problem is that the number of cores in NTT increases by powers of two, leaving no room for a middle solution.  

    \item Option 3: Use only one NTT unit in the RPAU and  reduce or completely hide the latency of the coefficient-wise operations by executing NTT and coefficient-wise polynomial operations in parallel.
    Using a few extra modular adder and multipliers, we can compute these cheap coefficient-wise operations in parallel to NTTs.    
    For example, using only four extra modular arithmetic cores, we can compute two dyadic polynomial arithmetic instructions (taking $2\times4,096$ cycles) concurrently to an NTT (taking around 7,168 cycles) and effectively reduce the latency of steps 7-8 in Algorithm~\ref{algo:relin_14} to the latency of one NTT only.  
\end{itemize}

We apply Option 3 as it is computationally fast and at the same time requires a minor increase in the logic area. The \texttt{RPAU.Dyadic} group in Fig.~\ref{fig:rpau} executes coefficient-wise instructions in parallel to NTT instructions in the \texttt{RPAU.All} group. We observe 40\% reduction in the latency at the price of only 20\% increase in the logic resources.  
Fig.~\ref{fig:mult_steps_parallel} gives a timing diagram and shows how we can speedup the computation of \texttt{HEAAN.Mult} and \texttt{HEAAN.Relin} using the two parallel compute groups. \color{black}
    
\subsubsection{Organization of On-chip Memory inside RPAU}~\label{sec:impl_mem}

\noindent\textbf{Peak memory requirement:} After optimizing the steps of homomorphic multiplication followed by relinearization, we observe that the peak memory requirement per RPAU is equal to storing seven residue polynomials (ciphertext-dependent data), and additional $2L$ residue polynomials of the key-switching-key. %
E.g., for Set-1 we have $L=7$ and therefore we need to store 21 residue polynomials on-chip for each RPAU to eliminate off-chip memory access completely during key-switching.
In a similar way, the peak memory requirement for Set-2 with $2N=2^{15}$ and $L=9$ becomes equivalent to storing 49 residue polynomials of degree $N=2^{14}$ per RPAU. To handle such large data, an option will be to use an off-chip memory. That will involve the cost of slow data exchanges. Note that the proposed parameter-flexible design methodology makes such a data-partitioning easy to implement. 

We take a step further and aim at keeping all the data in the on-chip memory for both Set-1 and 2. %
Neither BRAMs nor URAMs alone in U250 could store such a big data, but their combination with a few  tricks can.\\   

\noindent\textbf{On the fly key-switching key generation:} 
The large size of the key-switching key increases the peak memory requirement significantly. 
We notice that the $\mathtt{KSK}_0$ component of the key-switching key $\mathtt{KSK}$ is generated by expanding a pseudorandom public seed. The other component $\mathtt{KSK}_1$ is computed using the secret key. Therefore, the cloud can regenerate $\mathtt{KSK}_0$ itself from the seed. We use that to avoid storing half of the key-switching key by generating the pseudorandom polynomials of $\mathtt{KSK}_0$ on-the-fly using a PRNG in the hardware. Note that similar techniques are used in lattice-based post-quantum algorithms, example~\cite{saber_nist}, to generate their public polynomials. 

In our implementation, we use the 64-bit Trivium core~\cite{trivium_github}. The trivium core takes a 64-bit seed as input, performs 18 round initialization in 18 clock cycles, and thereafter  generates 64-bit random data per cycle without any stall. We use four Trivium cores per RPAU as there are four dyadic cores too per RPAU. These dyadic cores consume the Trivium$\times$4 generated pseudorandom words.

Overall, on the fly generation of $\mathtt{KSK}_0$ reduces the storage space for the key-switching key by half. For $2N=2^{15}$ and $L=9$ now we need to store only 31 $2^{14}$-degree residue polynomials in the on-chip memory per RPAU.\newline

\begin{figure}[t]
\centering
\includegraphics[width=5.2in]{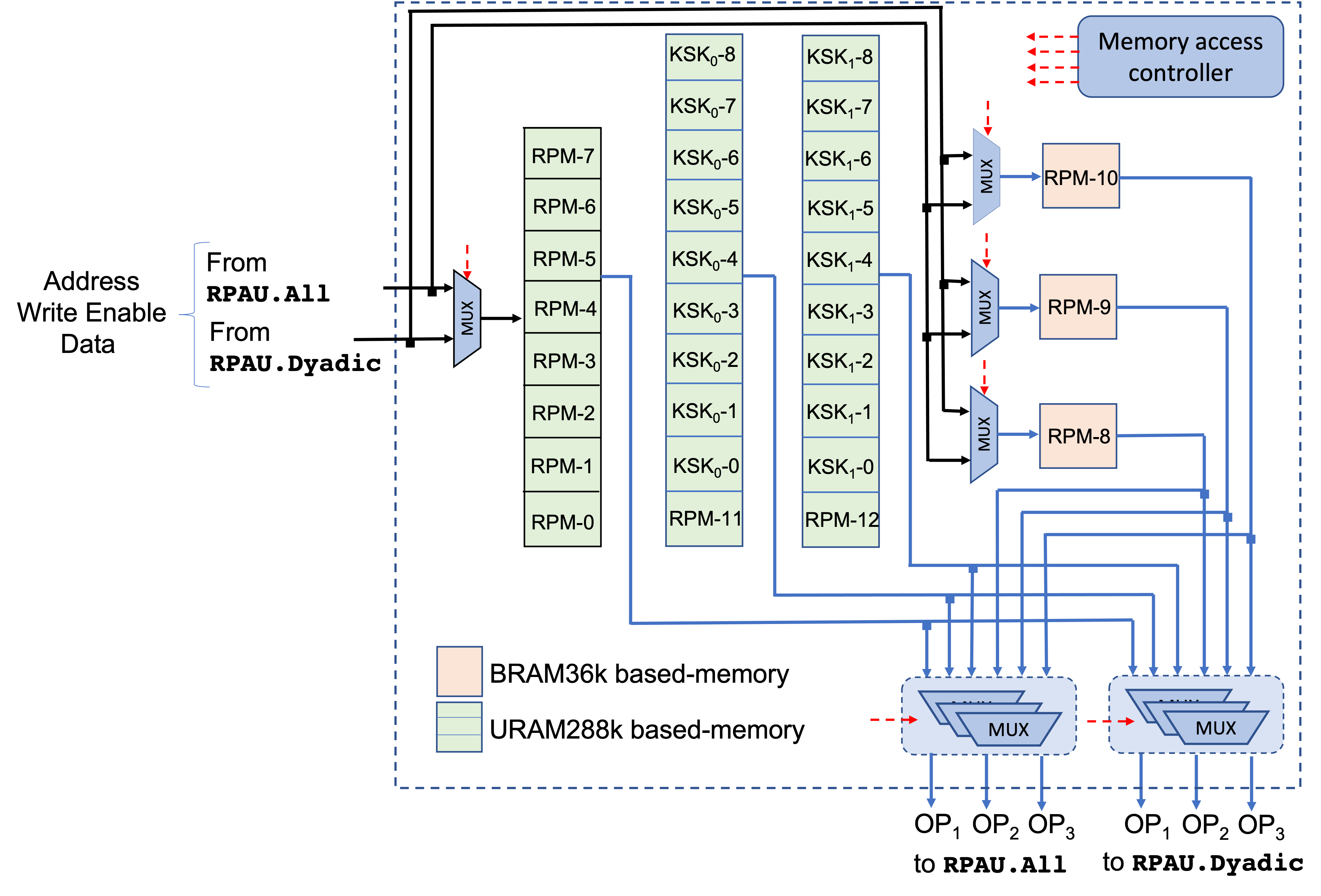}
\caption{Organization of memory elements inside the memory bank. One memory bank is connected to a one core of the NTT unit. There are 16 such memory banks inside each RPAU. Any residue polynomial is split into 16 fragments, and one fragment is stored in one RPM-$i$ of a memory module. }
\label{fig:RPAU_memory}
\end{figure}

\noindent\textbf{Memory unit:} 
We keep all the data variables for both the parameter sets Set-1 and 2 in the designed on-chip memory unit. We are the first to achieve fully on-chip computation of the key-switching operation for the two large parameters. 
On the target U250 FPGA, one URAM is 8 times larger than one BRAM36K. However, both types of memory have only two ports. 
Quantitatively, one $2^{14}$-degree residue polynomial needs 32 BRAM36K or 4 URAM slices. 
If a polynomial is stored using 4 URAMs, then due to URAM's I/O port limitation we cannot use all 16 cores of the NTT unit during an NTT/INTT operation. 
Hence, designing the memory unit of the RPAU requires careful consideration of the computation and architectural constraints. 

We make the memory organization modular inside the RPAU. To provide fast data access during NTT, we assign one `memory-bank' exclusively to each butterfly core. One `memory bank' is a heterogeneous collection of BRAMs and URAMs as shown in Fig.~\ref{fig:RPAU_memory}. 
The memory-bank of the $i$-th core of NTT keeps only the $i$-th fragments of all the 31 residue polynomials. 
In Fig.~\ref{fig:RPAU_memory} the abbreviation `RPM' stands for the residue polynomial memory. There are 13 RPMs as there are 13 ciphertext-dependent residue polynomials. 
Each RPM stores 1/16-th of the consecutive coefficients, i.e., 1,024 coefficients for $N=2^{14}$.

In the figure, we use the peach and light green colors to represent RPMs that are based on BRAMs and URAMs respectively. RPM-8, 9, and 10 are composed of BRAM36K slices and are physically separated. Hence, they can be read/written in parallel. Whereas, RPM-0-to-7 are implemented using a single pair of URAMs and are logically separated. Hence, only one of them can be read and only one of them can be written every cycle. 
Also, URAM-based RPM-11 and 12 are reserved for ciphertext-dependent residue polynomials.
Programmer decides which polynomial goes to which RPM taking care of data dependencies and access patterns of the HE subroutine.  
The `Memory access controller' block is responsible for handling memory accesses of the two parallel computing groups \texttt{RPAU.All} and \texttt{RPAU.Dyadic}. 
\color{black}

We also use URAMs to store key-switching key. We need to store $2L=18$ $2^{14}$-degree residue polynomials for key-switching key, which require 72 URAMs. Each address of a URAM can store 72-bit. For a 54-bit coefficient modulus, one $2^{14}$-degree residue polynomial can be stored in $(54 \cdot 4)/72=3$ URAMs instead of 4 by utilizing the left-over bits in each URAM address. Therefore, for RPAU units with  54-bit coefficient modulus, we only use 56 URAMs instead of 72 to store key-switching key.
Similarly, we use 63 URAMs instead of 72 for 60-bit coefficient modulus.

\subsubsection{Interconnecting Multiple RPAUs}\label{sec:rpau_ring}
The designed RPAU from the previous section is a fundamental compute element in our programmable architecture. Just using one RPAU in a time-shared manner we can process an HE subroutine. 
For fast computation time, we instantiate several RPAUs in parallel. As key-switching and re-scaling operations require exchanging the residue polynomials for different moduli, the RPAUs must perform data exchanges between them. If performed via a shared memory, such data exchanges will slow down the HE operations. On the other hand, connecting the RPAUs in the form of a star network will bring placement hurdles or may even make an actual HW implementation impossible on FPGAs. We present a novel way of connecting the RPAUs and optimally solve the problem. 

%

First, we introduce the challenges in interconnecting the RPAUs and then describe our solution. 
Like other large-scale Xilinx FPGAs~\cite{xilinx_ug872}, our Alveo U250 platform uses SSI technology and consists of four `semi-separated' SLR regions. Two neighbouring SLRs are connected using a limited number of wires. We observe that one SLR could fit up to three RPAUs and hence at least three SLRs are needed to implement Medha. Interconnecting the RPAUS must take SLR-to-SLR connection constraint into account to make implementation feasible.
Additionally, the communication unit between of the FPGA resides in SLR0 and SLR1. An input ciphertext should be sent efficiently and stored in the memory blocks of the RPAUs that reside in the other SLRs. Similarly, output polynomials should be read from different SLRs to SLR1.

The naive solution would be connect the RPAUs in a `Star' network keeping separate paths for each connection. We found that such an interconnection complicates the routing, bolts the number of nets crossing the SLRs, and ultimately reduces the clock frequency to around 50 MHz or less. 
%

%

\begin{figure}[t]
\centering
\includegraphics[width=3.8in]{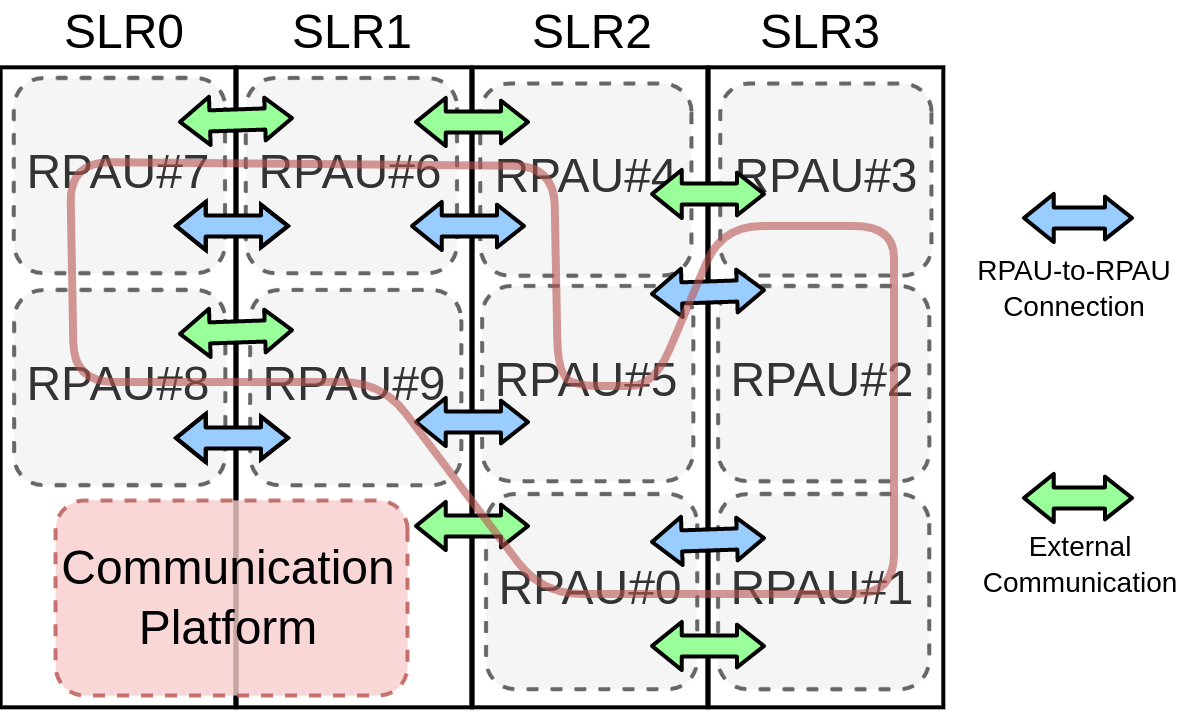}
\caption{The proposed 'ring' structured floorplan to minimize routing cost for the implementation with 10 RPAU units.}
\label{fig:ring_floorplan}
\end{figure}

We propose a 'Ring' interconnection of the RPAUs to reduce the routing: only two neighbouring RPAUs are connected. In many general-purpose applications, a ring network is considered slow due to its serial communication. Interestingly, after analyzing the computation steps of key-switching and re-scaling, we find that the exchange of residue polynomials between the RPAUs could be transformed into a \emph{broadcasting} protocol where only one RPAU broadcasts a polynomial at a time and all the remaining RPAUs receive. Such a transformation does not add any computation overhead. 
Therefore, in Medha we connect neighbouring RPAUs only and the arrangement of all the RPAUs looks like a ring. Each RPAU sends its data to any other RPAU through a chain of RPAUs.
In Fig.~\ref{fig:ring_floorplan}, we show placement of 10 RPAUs on the floor of the FPGA. The ring is marked with a red line. 
For external communication signals, we followed the 'ring' connection for sending data signals from SLR1 to other SLRs as shown in Fig.~\ref{fig:ring_floorplan}. 

\subsubsection{Program Execution Unit}

Our Medha is an instruction set architecture with its own program execution unit. An RPAU receives its instructions from a program execution unit. Using dedicated program controllers for each RPAU we can run asynchronously when there are no data dependencies between the RPAUs. However, the key-switching operation requires periodic data exchanges between RPAUs. Hence, we do not allocate dedicated program controllers for any RPAUs. By analyzing the computation steps in the homomorphic subroutines, we observe that most of the time the RPAUs could execute the same instruction in a SIMD manner. Only during the mod-down and rescaling operations, the program execution flow splits into two parallel branches: a subset of the RPAUs follow the first branch and the remaining RPAUs follow the second branch. Hence, Medha uses only two program controllers inside its program execution unit. We would also like to mention that by reducing the number of program controllers to two from `one for each RPAU' we greatly simplify the programming model of Medha.

\subsubsection{Scheduling of Homomorphic Operations}
The proposed Medha is an instruction-set architecture. Therefore, efficient translation of homomorphic operations into Medha instructions is crucial for achieving optimal performance results. 
In this section, we describe the scheduling of homomorphic operations for both parameter sets.
\\

\noindent\textbf{Scheduling for $N=2^{14}$:}
$\mathtt{HEAAN.Add}$ operation is straightforward and it can easily be implemented using two coefficient-wise addition instructions.
The $\mathtt{HEAAN.Mult}$ operation can be performed using coefficient-wise multiplication and addition instructions. 
As shown in Algorithm~\ref{algo:relin_14}, relinearization operation requires each residue polynomial in $\Tilde{\boldsymbol{d}}_{2}$ to be reduced by the other moduli (see step 7 of Algorithm~\ref{algo:relin_14}). Hence, all residue polynomials in $\Tilde{\boldsymbol{d}}_{2}$ first need to be transformed to the time-domain using one INTT instruction in the parallel RPAUs as shown in steps 1-3 of Algorithm~\ref{algo:relin_14}.
Since $\mathtt{HEAAN.Relin}$ operation needs $\Tilde{\boldsymbol{d}}_{2}$ output of $\mathtt{HEAAN.Mult}$ first, $\mathtt{HEAAN.Mult}$ and $\mathtt{HEAAN.Relin}$ operations can be performed efficiently in parallel. As soon as $\Tilde{\boldsymbol{d}}_{2}$ is generated, INTT operation is performed using the main computation core while remaining outputs of $\mathtt{HEAAN.Mult}$ can be generated using dyadic cores in parallel.
Then, the $i$-th sends the $i$-th residue polynomial in $\boldsymbol{d}_{2}$ to the other RPAUs through the ring structure. This operation is performed using an instruction that broadcasts a residue polynomial in an RPAU to other RPAUs.
The received $\boldsymbol{d}_{2}[j]$ are modulo-reduced by the other moduli and then converted back to the NTT domain. This procedure is repeated for $j=0,1,\ldots,l-1$ as shown in Algorithm~\ref{algo:relin_14}. Note that, the coefficient-wise multiplications and accumulations in the key-switching loop (steps 7-8 in Algorithm~\ref{algo:relin_14}) are performed in parallel to the NTTs using \texttt{RPAU.Dyadic}. 
Due to such parallel processing, the overall cycle count of $\mathtt{HE.Relin}$ is primarily determined by the latency of computing 1 INTT followed by $l-1$ NTTs. %
$\texttt{HEAAN.ModDown}$ and $\texttt{HEAAN.Rescale}$ operations are implemented sequentially using INTT, NTT and broadcasting instructions of the main core as they are not parallelizable in instruction-level. \newline

\noindent\textbf{Scheduling for $2N=2^{15}$:}
The scheduling of homomorphic operations for the parameter size $2N=2^{15}$ is very similar to the scheduling for the parameter size $N=2^{14}$ except for the routines involving \texttt{Split} and \texttt{Join} functions.
The \texttt{HEAAN.Add} and \texttt{HEAAN.Mult} operations can be performed efficiently using coefficient-wise addition and multiplication instructions on the split ciphertexts.
We first store all split ciphertexts $\mathtt{ct}^{(p)}$ and $\mathtt{ct}^{(b)}$ into the on-chip memory, then perform homomorphic operations $\mathtt{ct}^{(p)}$. After that, we move $\mathtt{ct}^{(m)}$ into the proper memory locations and repeat the same homomorphic operations. 

As shown in Algorithm~\ref{algo:relin_15}, the relinearization operation requires split/join and therefore $\mathtt{ct}^{(p)}$ and $\mathtt{ct}^{(m)}$ cannot be processed completely independently.  
As a consequence, \texttt{HEAAN.Relin} requires some extra on-chip data movement for split/join operations during the computations. Due to split/join and extra on-chip data movements, the homomorphic operations for the parameter size $2N=2^{15}$ costs more than $2\times$ clock cycles than the homomorphic operations for the parameter size $N=2^{14}$.  
\color{black}

\subsubsection{Hardware-Software Interfacing of the Overall System}

\begin{figure}[t]
    \centering
    \includegraphics[scale=0.90]{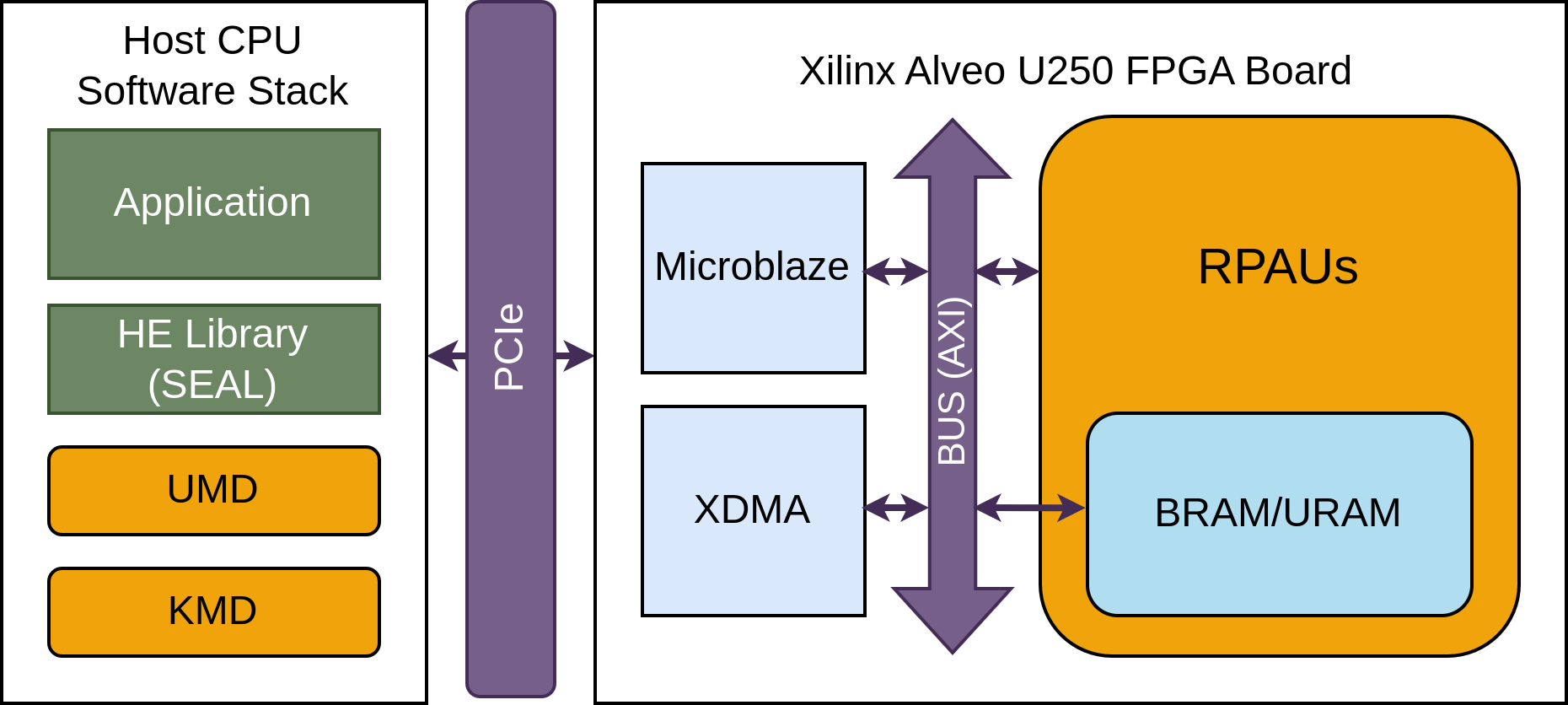}
    \caption{CPU-FPGA interface and software stack}
    \label{fig:cpu_fpga_stack}
\end{figure}

We implemented a proof-of-concept software stack (Fig.~\ref{fig:cpu_fpga_stack}) consisting of a SEAL library, User-Mode Driver (UMD), and Kernel Mode Driver (KMD). The UMD provides an interface layer for SEAL, and KMD supports the scheduling of jobs. When a SEAL command (supported by Medha) is executed, the corresponding UMD-API is called to submit the command with the required parameters to KMD's job queue as a job. Next, KMD's job scheduler sends the job to Medha. When Medha completes its task, the result is read through the PCIe interface. All data communications are performed using XDMA~\cite{xilinx_xdma} for fast transfers. We use the MicroBlaze (Xilinx's microprocessor) unit for controlling the communication between the host CPU and the RPAUs, and also for monitoring the entire FPGA system. 

%% file: 6.results.tex
\section{Results}\label{sec:results}

We described Medha in Verilog HDL and implemented in Xilinx Alveo U250 card. We used Vivado 2019.1 tool with a performance-optimized implementation strategy.
The implementation of Medha employs 10 RPAU units and works for the Set-1 and Set-2 from Table~\ref{tab:results_paramset}.

\subsection{Timing Results}

The proposed implementation runs at 200MHz.
In Table~\ref{tab:results_timing}, we present the cycle count, latency (in $\mu$ sec) and throughput results for each low-level instruction and high-level operations for the Set-1 and Set-2.
The cycle counts were collected using a hardware-based counter. 
The low-level instructions for synchronizing main/dyadic cores, synchronizing the program controllers, and ending the program do not consume any clock cycles, thus they are not included in Table~\ref{tab:results_timing}.
Since we use only on-chip memory (i.e. registers and BRAMs/URAMs) during the computations, the proposed architectures do not have any DDR data transfer overhead.
\begin{table}[t!]
\caption{Performance of each Instruction/Operation}\label{tab:results_timing}
\centering
\begin{tabular}{c | l | r r r}
\hline

\multirow{2}{*}{\textbf{Set}}&\multirow{2}{*}{\textbf{Instruction/Operation}}  & \multicolumn{1}{c}{\textbf{Clock}}  & \multicolumn{1}{c}{\textbf{Lat.}}  & \multicolumn{1}{c}{\textbf{Throug.}} \\ 
&                            & \multicolumn{1}{c}{\textbf{Cycles$^*$}} &   \multicolumn{1}{c}{($\mu$ sec)}   & \multicolumn{1}{c}{(per sec)}\\  \hline\hline

      &$N$-pt NTT             & $\approx$ 7,200 & 36 & 27,777 \\ 
      &$N$-pt INTT            & $\approx$ 7,200 & 36 & 27,777 \\
Set-1 &RPAU-to-RPAU broadcast & $\approx$ 512 & 2.56 & 390,625 \\
\&    &Scale by $q_i^{-1} \pmod{q_j}$ & $\approx$ 512 & 2.56 & 390,625 \\
Set-2 &C.wise Add/Sub/Mult (main)    & $\approx$ 512 & 2.56 & 390,625 \\ 
      &C.wise Add/Sub/Mult/MAC (dyd) & $\approx$ 4,096  & 20.48 & 48,828 \\ 
      &Split/Join & $\approx$ 1,024  & 5.12 & 195,312 \\
      &Automorphism & $\approx$ 512  & 2.56 & 390,625\\ \hline

      & Hom. Add                   & 1,152    & 5.76   & 173,611\\ 
Set-1 & Hom. Mult. + Relin.        & 99,448  & 497.24 & 2,011 \\ 
      & Rescale                    & 34,430       & 172.15     & 5,808    \\ 
\hline

      & Hom. Add                   & 2,865  & 14.33 & 69,783 \\ 
Set-2 & Hom. Mult. + Relin.        & 274,885  & 1,374.42 & 727 \\
      & Rescale                    & 75,464  & 377.32 & 2,650 \\ 
\hline

\hline

\end{tabular}
\begin{flushleft}
$^*$All cycle counts are reported using a hardware-based counter.
\end{flushleft}
\end{table}

\subsection{Resource Utilization Results}
Table~\ref{tab:results_7rpau} shows the resource requirements for the complete processor architecture, one  RPAU (with 54-bit and 60-bit modulus), one butterfly core, and one twiddle factor generation core.  
The area figures for the complete processor also include the ‘Platform’ unit that is responsible for the communication between the FPGA and the host CPU. 
The overall design consumes 63.2\% of LUTs, 23.2\% of registers, 58.6\% of BRAMs, 72.7\% of URAMs and 29.3\% of DSPs available on the Alveo U250 Card.
%
Since we target performing homomorphic operations using only on-chip memory, the proposed design has very high BRAM and URAM utilization.
Although the resulting URAM utilization allows instantiating more RPAU units, there are several reasons preventing us from adding more RPAU units.
Firstly, the 'Platform' unit consumes more resources than one RPAU unit and it resides across two SLR regions, which limits the placement of new RPAUs.  
Secondly, increasing the number of RPAU units (hence the number of RNS bases) will increase the size of key-switching key. Therefore, each RPAU unit will employ more URAMs to store keys. Since each SLR region has only 320 URAMs in Alveo U250 Card, placing 3 RPAUs into a single SLR region will be much harder.
Also, high routing cost of SLR-to-SLR communication prevents placing one RPAU across multiple SLR regions. 
Finally, adding a new RPAU complicates the routing and decreases the maximum achievable clock frequency of the overall design.
Thus, we set the number of RPAU units as 10.

Alveo U250 FPGA card can have a 215W maximum power consumption level~\cite{xilinx_alveo_datasheet} with a typical consumption level of around 100W. 
For Medha, Vivado reported on-chip power consumption of 62.12W.

\begin{table}[t!]
\caption{Resource Utilization of Arithmetic Modules on Alveo U250 Card}\label{tab:results_7rpau}
\centering
\begin{tabular}{l | r r r r r}
\hline
\textbf{Modules} & \textbf{LUTs} & \textbf{REGs} & \textbf{BRAMs} & \textbf{URAMs} & \textbf{DSPs}  \\ \hline\hline

Processor              & 1,093,250 & 803,713 & 1,576.5 & 931 & 3,607 \\ 
$\lfloor$Platform          & 128,237 & 132,090 & 296.5 & 1 & 7 \\
$\lfloor$Prog. Controller  & 860     & 276 & --    & -- & -- \\
$\lfloor$Cryptoprocessor              & 963,131 & 669,353 & 1,280    & 930 & 3,600 \\ \hline

RPAU Unit (54-bit)         & 76,474 & 57,180 & 128 & 92 & 360 \\ 
$\lfloor$Memory Core   & 35,369 & 8,045   & 96 & 92 & -- \\
$\lfloor\mathtt{KSK}_0$ Core   & 1,934  & 1,654  & -- & -- & -- \\
$\lfloor$Dyadic Core   & 6,872  & 3,571  & -- & -- & 40 \\
$\lfloor$Main Core     & 32,214 & 41,248 & 32 & -- & 320 \\ \hline

RPAU Unit (60-bit)         & 79,631 & 57,479 & 128 & 102 & 360 \\ 
$\lfloor$Memory Core   & 35,128 & 7,352   & 96 & 102 & -- \\
$\lfloor\mathtt{KSK}_0$ Core   & 2,031  & 1,491  & -- & -- & -- \\
$\lfloor$Dyadic Core   & 6,902  & 3,700  & -- & -- & 40 \\
$\lfloor$Main Core     & 35,488 & 42,324 & 32 & -- & 320 \\ \hline

Butterfly Unit         & 1,358  & 1,625   & --  & -- & 10 \\ 
$\lfloor$Modular Mult. & 533   & 782   & --  & -- & 10 \\ \hline

TF Gen. Unit           & 755  & 962  & 2   & -- & 10 \\ \hline

\end{tabular}
\end{table}

\subsection{Comparison with Related Works}

\noindent\textbf{Comparison with SEAL:} There are various highly-optimized software implementations of the HEAAN scheme based on homomorphic encryption libraries such as Microsoft SEAL~\cite{SEAL3.6} and Palisade~\cite{PALISADE}. We compare the performance of Medha with the single-threaded software implementation of the RNS-HEAAN on highly-optimized homomorphic encryption library Microsoft SEAL~\texttt{v3.6}~\cite{SEAL3.6}. To present a fair comparison, we modified the SEAL accordingly to work with the parameter sets defined in Table~\ref{tab:results_paramset}. 
The latency of high-level homomorphic operations in SEAL~\cite{SEAL3.6} and its comparison to Medha for Set-1 and Set-2 are presented in Table~\ref{tab:results_seal}. 
The timing results of SEAL are obtained on an Intel i5-6200U CPU @ 2.30GHz $\times$ 4 with 16 GB RAM using \texttt{gcc} version \textsc{9.3} in Ubuntu \textsc{20.04.2} LTS. 
The proposed architectures with Set-1 and Set-2 showed 68.06$\times$ and 78.87$\times$ performance improvements, respectively, for the homomorphic multiplication with relinearization operation compared to the SEAL-based implementation. The effectiveness of our approach increases with the larger parameter sets.
In Table~\ref{tab:results_seal}, we also provide the performance results of SEAL from a single-threaded Intel Xeon(R) Silver 4108 running at 1.80 GHz, which was used in the HEAX paper~\cite{heax}, for the Set-1 parameter. Compared to this result, Medha shows 134.07$\times$ speed-up for homomorphic multiplication and relinearization operations.\\

\begin{table}[t!]
\caption{Latency Comparison with the SEAL~\cite{SEAL3.6} and HEAX~\cite{heax}}\label{tab:results_seal}
\centering
\begin{tabular}{c | l | r r r}
\hline

\multicolumn{1}{c|}{\textbf{Set}}& \multicolumn{1}{c|}{\textbf{Work}} & \multicolumn{1}{c}{\textbf{Add}} & \multicolumn{1}{c}{\textbf{Mult. + Relin.}} & \multicolumn{1}{c}{\textbf{Rescale}} \\\hline\hline

%

\multirow{7}{*}{Set-1} & \textbf{Medha}               & 5.76 $\mu s$  & 497.24 $\mu s$ & 172.15 $\mu s$ \\ \cline{2-5} 
 
 & \multirow{2}{*}{SEAL~\cite{SEAL3.6}}   & 418  $\mu s$  & 33,844 $\mu s$ & 6,429 $\mu s$ \\
 &                                        & (\textbf{72.56$\times$}) & (\textbf{68.06$\times$}) & (\textbf{37.34$\times$})  \\

 & \multirow{2}{*}{SEAL~\cite{heax}}   & --   & 66,666 $\mu s$ & -- \\
 &                                        & -- & (\textbf{134.07$\times$}) & --  \\
 
 & \multirow{2}{*}{HEAX~\cite{heax}}   & --  & 1,182.27 $\mu s$ & -- \\
 &                                        & -- & (\textbf{2.37 $\times$}) & --  \\
 
\hline

\multirow{3}{*}{Set-2} & \textbf{Medha}              & 14.33 $\mu s$  & 1,374.42 $\mu s$ & 377.32 $\mu s$ \\ \cline{2-5} 
 & \multirow{2}{*}{SEAL~\cite{SEAL3.6}}   & 921  $\mu s$  & 108,401 $\mu s$ & 16,438 $\mu s$ \\
 &                                        & (\textbf{65.03$\times$}) & (\textbf{78.87$\times$}) & (\textbf{46.56$\times$})  \\
\hline

\end{tabular}
\end{table}

\begin{table}
    \centering
    \begin{scriptsize}
    \begin{tabular}{l|c|c|c|c|c|c}
    \hline
        \multirow{3}{*}{\textbf{Work}} & {\textbf{Platform}} &\textbf{Real} & \textbf{Best} & \textbf{Freq.} & \textbf{Area}$^{\S}$ & \textbf{Plat.}$^{a}$\\
                                       & \textbf{(A:ASIC)}          & \textbf{HW?}       &  \textbf{speedup}           &  \textbf{(\textit{GHz})}  & (\textit{mm$^2$} \textbf{for ASIC})  & \textbf{price} \\
                                       &            &           &  \textbf{w.r.t. SW}                  &                         & (\% \textbf{for FPGA}) & \textbf{(US\$)} \\ \hline

       F1 \cite{feldmann_2021f1}       & A-14/12nm   & $\times$ &  17K$^{*}$  & 1 to 2 &  151.4  & >4M \\
       BTS \cite{BTS_isca}             & A-7nm    & $\times$ &  2.2K$^{*}$         & up to 1.2 &  373.6  & >10M \\
       BASALISC \cite{basalisc}        & A-12nm   & $\times$ &  4K$^{*}$           & 1 to 2 &  150 & >4M \\
       CraterLake \cite{CraterLake}    & A-14/12nm   & $\times$ &  8.6K$^{*}$  & 1 to 2 &  472.3  & >13M\\
       ARK \cite{ARK}                  & A-7nm    & $\times$ &  36K$^{*}$     & NA &  418.3  & >11.5M\\
       CoFHEE~\cite{cofhee}            & A-55nm   & \checkmark   & 2.5       & 0.25     &  15   & >55K \\\hline
       
       ReMCA~\cite{ReMCA}                 & Virtex-7   & $\times$   & NA        & 0.25     &  6.5L + 20.8B    &26K \\
       \cite{roy_hpca}                 & ZCU102   & \checkmark   & 13        & 0.20     &  50L + 90B    &3.3K \\
       HEAWS \cite{he_aws}             & AWS-F1   & \checkmark   & 20         & 0.25     &  75L + 83B    & Rent  \\
       HEAX \cite{he_aws}              & Stratix 10   & \checkmark   & 164$^{\dag}$     & 0.30   &  64L + 80B    & 8.5K \\
       Medha$^{\ddag}$                           & U250     & \checkmark   & 137$^{\delta}$     & 0.20     &  55L + 72B    & 9K \\\hline         

    \end{tabular}
    \caption{Comparison of Hardware Accelerators for Homomorphic Evaluation}
    \label{tab:comp_all_hardware}
    \end{scriptsize}
    \begin{flushleft}
    {\footnotesize
    $^*$: Throughput is estimated by simulating a model of the accelerator.  \\
    $^{\dag}$: Best asymptotic throughput from hardware.  \\
    $^{\delta}$: Throughput calculated from latency of one operation.   \\
    $^{\ddag}$: Latency of HEAX is 2.37$\times$ slower than Medha. \\
    $^{\S}$: In the Area column, `L' and `B' stand for \% of logic and on-chip RAM elements used. \\
    $^{a}$: In the Price column, the estimated ASIC fabrication costs are based on~\cite{fab_pricing}.}
    \end{flushleft}
\end{table}

\noindent\textbf{Comparison with HEAX:} %
The fairest comparison is with the HEAX processor~\cite{heax}.
It is the only prior art for actual FPGA-based implementation of RNS-HEAAN and with the same parameter set (Set-1). HEAX and Medha follow significantly different design methodologies. Unlike Medha, HEAX unrolls the key-switching of RNS-HEAAN into steps and then instantiates one dedicated block per step to attain high throughput. These blocks are cascaded to realize a block-pipeline architecture. There are a total of six block-pipeline stages in the implementation of the key-switching operation. During a key switching, all the residue polynomials are processed one-by-one through the pipeline stages. Thanks to such unrolled and block-pipelined architecture, HEAX achieves a very high asymptotic throughput of 2,616 homomorphic multiplication including key-switching operations per second at 300MHz on a Stratix10 FPGA for the Set-1 parameter.  

In comparison, Medha is an instruction-set architecture with programmability, and it reuses the RPAUs again and again for computing different steps of various homomorphic routines. 
Naturally, Medha is a low latency-oriented architecture. It still achieves a competitive throughput (i.e., time/latency of one operation) of 2,011 homomorphic multiplications including key-switching operations per second while running at a lower clock frequency of 200MHz for the Set-1 parameter. 

Latency-wise, Medha is more than 2$\times$ faster than HEAX as shown in Table~\ref{tab:results_seal} for the Set-1 parameter. As the latency figures of HEAX are not specified in~\cite{heax}, we estimate them based on the computation flow diagram from Table~5 and Figure~6 of~\cite{heax} as follows. There are six stages of block-pipeline processing during a key-switching (the last row or Set-C of Table-5 in~\cite{heax}) and the stages have been designed to have similar cycle counts. The first stage uses an 8-core inverse-NTT with at least 14,336 cycles latency. 
Thus, each stage has roughly 14,336 cycles of latency. 
As there are seven RNS-moduli and 18 pipeline stages including a one-core INTT stage with 114,688 cycles latency (see Figure~6 of~\cite{heax}), computing a full key-switching will take at least 358,400 cycles. In comparison, our Medha has a latency of 99,448 cycles only for computing one homomorphic multiplication plus a key-switching for the Set-1 parameter.

From an architect's perspective, Medha has five main advantages over HEAX: $(i)$ Medha shows better latency performance with a competitive throughput, $(ii)$ Medha uses only on-chip memory during the computations while HEAX needs off-chip memory communication during the key switching operation for Set-1, $(iii)$ HEAX uses a fixed-pipelined architecture tailored for HEAAN scheme while Medha's instruction-based architecture allows flexibility, $(iv)$ HEAX uses separate arithmetic units to perform homomorphic multiplication and relinearization, and it does not perform rescaling, homomorphic addition/subtraction, and rotation operations in hardware (see Fig. 7 in~\cite{heax}), and $(v)$ HEAX provides support for polynomial degree $2^{14}$ while Medha supports homomorphic operations for multiple parameters with $N=2^{14}$ and $2N=2^{15}$. 
We should also note that the design principles of Medha were not selected to outperform HEAX. Our main goal was to gain fast computation time while keeping programmability. Our results show that our instruction-set architecture can outperform a block-pipelined and specific architecture in terms of latency. Our results bring a new direction to the design space, otherwise HEAX's one order superiority over the previous programmable processors~\cite{roy_tc2018, roy_hpca} would have indicated that specific and block-pipelined processors are the way to accelerate HE in HW.   
Thanks to the significantly lower latency, Medha would be advantageous for practical homomorphic applications compared to HEAX. The asymptotic throughput of HEAX is achievable only if we assume that in the application there are plenty of data-independent homomorphic operations most of the time and that there is no overhead at the host side (e.g., a SW system) concerning managing the input-output ciphertexts. 
Note that, already due to the batching of messages, a homomorphic operation for the Set-1 parameter implicitly performs the operation on $N/2 = 8,192$ slots in parallel. 
In real-life applications, there will be data dependencies and additionally, a SW host (which is running the application and using the HW as a service) will introduce some overhead in the processing of operand and result. For example, consider computing the $e^5$ from an encryption of $e$. Hence, the overall processing time of an application will greatly be determined by the latency instead of throughput.

There is one more advantage of using a low-latency system from a full-stack implementation point of view. Different homomorphic compilers have been designed to translate plaintext applications into homomorphic applications automatically. These compilers try to reduce execution time by reducing the number of homomorphic multiplications and depth of multiplication chains. If the latency is used as a `cost'-metric, then the optimization task for a homomorphic compiler becomes simpler. On the other hand, if the accelerator is throughput-oriented, then the tasks for a homomorphic compiler become more challenging as it has to identify different ways of parallelization and also make necessary arrangements for handling the parallel ciphertexts (which are large in size).\\

%

\noindent\textbf{Comparison with F1:} \cite{feldmann_2021f1} proposed 
an instruction-based wide-vector processor architecture `F1' and presented simulation-based performance estimates using a cycle accurate C/C++ model of the hardware. In 14nm/12nm process, logic synthesis of F1's architecture description reported an absolute area of 151 mm$^2$ and 1 GHz operating frequency. 
F1 has been optimized for throughput. The authors estimate that F1 will achieve a throughput of 500,000 homomorphic multiplication and relinearization per second at 1 GHz for Set-1 parameter. How the throughput estimate is calculated is not discussed in the paper. Although the paper mentions that F1 decouples data movements from computation, it does not mention if the throughput estimates assume that the external memory device gives the highest data rate. Benchmarking of different applications on conventional CPUs show that external memory devices do not operate at their commercially advertised peak data rate. 
Latency figures for homomorphic operations are not presented in the paper~\cite{feldmann_2021f1}.

We present a normalized performance comparison between F1 and Medha. F1 has 16 high-level clusters, each having one `unrolled’ NTT and many dyadic units. The unrolled NTT uses a condensed 2D array of butterfly circuits. 
The number of multipliers in one cluster of F1 is roughly 50\% of the total multipliers available in one Alveo U250 board. 
If we assume that the FPGA keeps one cluster of F1 and runs at 200 MHz, then the performance of F1 will drop by $80$ times (16 times due to the use of only one cluster and another 5 times due to slower frequency), excluding the overhead of off-chip access. In this scenario, F1 and Medha will have almost the same speed. If the overhead of off-chip access is considered, then F1 will be slower than Medha. In reality, F1 may not be implementable in present-day FPGAs. Unrolled NTT was first proposed in~\cite{ring_lwe_ches2012} and was found to be impractical in FPGAs. As the multipliers are distributed homogeneously on the FPGA floor, and SLR-to-SLR connections are limited in number, it is likely that F1's unrolled NTT may not fit in FPGAs. The authors of F1 did not report any FPGA-based results. 

There are several additional limitations of the F1 architecture. Firstly, F1 uses 32-bit moduli whereas Medha uses 60 and 54-bit moduli similar to SEAL. Thus F1 supports lower-precision arithmetic of encrypted real numbers than Medha and SEAL. With 32-bit moduli, Medha could accommodate around 16 RPAUs and therefore support around 15 multiplicative levels for $N=2^{14}$. 
Secondly, unlike Medha's efficient ring structure, F1 uses a crossbar network for realizing cluster-to-cluster communication. The proposed crossbar network uses a 10 mm$^2$ chip area with 19.6W design power. 
The Medha's ring structure limits the direct communication of an RPAU to only its two neighboring RPAUs, hence it reduces the number of interconnects significantly. Although F1 supports multiple polynomial degrees, it provides support only up to polynomial degree $2^{14}$ while our Medha supports the polynomial degrees $2^{14}$ and $2^{15}$.

Note that there is no real ASIC hardware of F1.
Its simulation-based performance estimates will be impressive only if a real F1 chip is ever built. Authors of F1 consider chip-simulation as future work. 
Newer architectures BTS~\cite{BTS_isca}, CraterLake~\cite{CraterLake} and ARK~\cite{ARK}, BASALISC~\cite{basalisc} claim further performance improvements by designing larger processors. Similar to F1, they present performance and area estimates based on simulation and logic synthesis.
Table~\ref{tab:comp_all_hardware} compares them and other hardware accelerators with Medha.\\
%

%

\noindent\textbf{Comparisons with other HW implementations:} The works in~\cite{roy_tc2018,roy_hpca,he_aws} present the FPGA implementations for the high-level operations of the BFV scheme. In~\cite{roy_tc2018}, the authors proposed an implementation targeting vert large parameter set (namely $N=2^{15}$ and $\log_2Q = 1228$) and their implementation suffers from off-chip memory communication requirements. 
The works in~\cite{roy_hpca} and~\cite{he_aws} use smaller parameters (namely $N=2^{12}$ and $\log_2Q = 180$) and shows performance improvements for homomorphic multiplication operation compared to the FV-NFLlib. 
Our design shows better performance and supports significantly larger parameter set.

%
%
%

There are simulation-based works targeting acceleration of the BFV scheme using \textit{compute-in-memory} approach, where computations are performed using arithmetic units very close to the memory elements~\cite{Reis_2020,takeshita2020}. 
As these works target a significantly different platform, presenting a fair comparison between these works and Medha is not feasible.\\
%
%

%

\noindent\textbf{Comparisons with GPU implementations:} To the best of our knowledge, there are only two GPU implementations for the RNS-HEAAN scheme in the literature~\cite{JKACL21,badawi20}. 
For the parameter set $N=2^{14}$ and $\log_2Q=360$, \cite{badawi20} performs homomorphic addition and homomorphic multiplication with relinearization in 0.04 ms and 0.74 ms, respectively. For a similar parameter set (Set-1), our architecture shows 6.9$\times$ and 1.5$\times$ better performance compared to their system running on an NVIDIA DGX-1 multi-GPU system with 8 V100 GPUs for homomorphic addition and homomorphic multiplication with relinearization, respectively.
Also, their multi-GPU platform NVIDIA DGX-1 has a reported maximum power consumption level of 3,500W which is 16$\times$ higher compared to the Alveo U250 board (at peak power).
The NVIDIA DGX-1 platform costs \$49,000 which is 6.3$\times$ higher than the Alveo U250 board.
Therefore, Medha running on an Alveo U250 board would be a more power-efficient and cheaper accelerator solution for homomorphic applications.

The work in~\cite{JKACL21} supports bootstrapping and focuses on memory-centric optimizations for an NVIDIA Tesla V100 GPU.
Their work targets very large parameter set, namely $N=\{2^{16},2^{17}\}$ and $\log_2Q \approx 2300$. Therefore, it is not easy to perform a fair comparison between their work and our architecture. 
There are also other GPU-based accelerator implementations in the literature targeting other HE schemes (i.e., BFV)~\cite{cuhe2015,badawi18,badawi19,badawi20_2} or partial operations such as NTT~\cite{zhai2021accelerating,ozerk2022efficient}. \\
%

%

\noindent\textbf{Comparisons with CHET and Cheetah:}
CHET~\cite{chet} proposes a compiler that selects the optimum parameters for HE applications and uses SEAL for performing the evaluations. The comparison of Medha with SEAL shows more than 2-orders of magnitude improvement. Cheetah~\cite{cheetah} focuses on optimizing homomorphic deep neural network applications using parameter tuning and operator scheduling at the application level. 
The paper presents area estimation of a conceptual hardware accelerator depending on parameters.
Actual hardware implementation is not provided. Their conceptual hardware uses 545 mm$^2$ area in 5nm technology with a 400 MHz clock frequency. 
Individual performance of neither the fundamental homomorphic routines (e.g., multiplication, key-switching) nor building blocks (e.g., NTT, dyadic-mult) are provided. So it is hard to compare Medha with Cheetah. 

Medha focuses on the acceleration of homomorphic operations and evaluations using a hardware platform, while CHET and Cheetah target optimum evaluations of HE applications by efficient automation. Medha’s existing FPGA implementation chooses the maximum parameter set that can be accommodated in the Alveo U250 board. Unlike other papers targeting neural-network applications, Medha provides general accelerator design concepts for HE that can be applied to any application when a sufficiently large platform becomes available.

\noindent\textbf{Benchmarking of homomorphic applications:}
The current FPGA implementation of Medha can be used to run any homomorphic application that requires a parameter set within Set-2 of Table~\ref{sec:params}. %
Medha can homomorphically evaluate inferences using low latency machine learning models like Logistic Regression~\cite{basalisc}, LeNet-5~\cite{chet}, (low latency cryptonets) LoLa-CIFAR and LoLa-MNIST~\cite{feldmann_2021f1}, etc.\cite{badawi20,gmdh},  that require parameters within Set-2. %
Apart from evaluating those models, our parameter set is sufficient for approximate computation of various mathematical functions like logistic, exponential, sin/cos/tan, etc., as shown in~\cite{approx}. 

We present \textcolor{black}{end-to-end application-benchmark results obtained from the hardware design of Medha} for evaluating the award winning logistic regression model~\cite{KSKLC18} of iDASH2017 competition. 
The model gives the probability of cancer and is trained on 1,579 samples and 18 features. We expand and evaluate the logistic regression function until $x^7$, which is sufficient for the application. 
Evaluating the logistic regression model requires 7 rotation, 11 rescaling, 5 multiplication and relinearization operations. The regression model is implemented using the parameter set $N=2^{14}$ and $\log_2(pQ) = 384$ in both SEAL and Medha. 
SEAL finishes one batch of inference, which encodes 455 samples with 18 features, in 0.7 seconds. In comparison, Medha takes only \textcolor{black}{10.85} ms on average, including 4.25 ms for communication. 
\textcolor{black}{Performing one batch of logistic regression inference in Medha takes around 1.3M clock cycles with 834 low-level instructions.}
\textcolor{black}{Note that the cost of sending evaluation and rotation keys to the Medha is not included in the end-to-end latency since they need to be sent only once in the beginning.}
Compared to SEAL, Medha evaluates the model around \textcolor{black}{64$\times$}  faster for this application benchmarking. 
The communication between CPU and FPGA happens using XDMA. Our experiments show that the cost of sending one polynomial of degree $2^{14}$ from host CPU to the FPGA using XDMA is taking 125$\mu s$ on average. Medha uses its on-chip memory to store the key-switching keys and avoids off-chip communication during the logistic regression operation.

PrivFT implementation~\cite{badawi20} proposed for text classification using homomorphic encryption runs a shallow network consisting of an embedding (hidden) layer and an output fully connected layer. For the parameter $N=2^{14}$ and $\log_2(pQ) = 360$ (i.e., $L = 5$), PrivFT inference takes a ciphertext and performs three multiplications with scalars, 61+14=75 additions, and 14 rotations. %
Medha with Set-1 can perform the same operation in 0.36 seconds and achieve $1.8\times$ speedup over the GPU implementation of PrivFT~\cite{badawi20}.

In \cite{gmdh}, the privacy-preserving forecasting application in the smart grid scenario uses a GMDH network having three hidden layers. It is evaluated for a smaller parameter (namely $N=2^{12}$ and  $\log_2Q = 186$ with $L=3$). Medha's Set-1 parameter is overly larger than theirs, yet Medha performs this evaluation $150\times$ faster. Every half an hour, Medha can evaluate the energy price forecast for $60,000$ apartments, which is sufficient for a small city, thus making Medha useful in the real-world scenario.\\ 
%

%

The state-of-the-art FPGA-based accelerator~\cite{heax} HEAX did not report any application benchmark. We report a  significant limitation of HEAX when application benchmark is taken into account. As stated earlier, the HEAX architecture keeps each major polynomial arithmetic step as a dedicated block. The blocks are cascaded one after another to realize the block pipelined end-to-end architecture of HEAX. In such a system, the host computer must feed the FPGA with operand residue polynomials as fast as the compute-stages in the pipelined architecture -- otherwise there will be bubbles in the pipeline stages. %
Our experimental results show that the CPU-FPGA XDMA communication is slow enough to degrade the performance of HEAX significantly. Furthermore, HEAX cannot take advantage of data locality in applications as any computed result (i.e., residue polynomial) must come back to the CPU before it can be processed again by the FPGA. In comparison, Medha is able to exploit data locality in the logistic regression application.

ASIC accelerators such as F1~\cite{feldmann_2021f1}, BTS~\cite{BTS_isca}, CraterLake~\cite{CraterLake}, etc.
target applications that require bootstrapping. They claim that their accelerators `compute' deep learning applications several thousands of times faster than software. In reality, such speedup claims are conjectures as no real ASIC prototypes of these accelerators yet exist. F1~\cite{feldmann_2021f1} and CraterLake~\cite{CraterLake} obtained benchmark estimates using software models of their architectures. Sec.~\ref{sec:related_works} discussed several bottlenecks that come in the way of obtaining real chips from such large architectures. Furthermore, the performance estimates assume that different processing stages in their architectures offer their maximum throughput. E.g.,~\cite{CraterLake} assumes a high data rate of 512 GB/s for external memory. As real-life applications typically do not see the maximum throughput of main memory, lower speedups will be observed when homomorphic applications are run.

%% file: 7.conclusion.tex
\section{Discussions}\label{sec:discuss}

\noindent\textbf{The hardest implementation challenge that we faced:}
We made a real hardware accelerator for homomorphic encryption after overcoming several difficult design challenges during different steps of the implementation flow. %
The hardest challenge we faced during the implementation was finding an efficient placement of the proposed design on an actual FPGA board. 
In a large design, although individual building blocks achieve high clock frequency, the overall design still can end up with a low clock frequency due to congestion issues (or even become impossible to route). Manual placement of building blocks is needed to realize an efficient implementation. It took several months of experiments and many design iterations for us to find an efficient placement of RPAU units and a way to move data efficiently across a large FPGA, which yields high clock frequency.
The first step was detecting problematic timing paths which happened to be the data signals for external communication routing from SLR1 to other SLR regions and placing extra pipeline buffers in these paths for providing flexibility in routing and improving the clock frequency. However, these additional buffers showed little or no improvements for the clock frequency due to the inefficient placement of RPAU units. 
Finally, the `ring-like' interconnection detailed in Sec.~\ref{sec:rpau_ring}, where each RPAU is constrained to a certain SLR region, made it possible to achieve a high clock frequency. These real engineering challenges do not appear when designers go up to the logic-synthesis of an architecture and report simulation-based performance estimates. \\

\noindent\textbf{Performance/cost estimate for larger parameters:}
In this paper, we implemented Set-1 and 2 and both implementations do not require off-chip communications during a key-switching. For larger parameter sets, it will no longer be possible to avoid off-chip data transfers. 
In such implementations, the proposed flexible design method will be particularly useful as it enables the re-use of smaller arithmetic units and memory. 
Now, we present estimates for larger parameter sets assuming that the resources of the target FPGA scale appropriately.
For $N=2^{16}$ and $L=9$, the peak memory requirement for the ciphertext-dependent data and key-switching key will be 25 and $4L=36$ residue polynomials, respectively. Following the same practices from Sec.~\ref{sec:impl_mem}, each RPAU will require 96 BRAMs and 208 URAMs. Then, the overall design will consume more than $2\times$ more URAMs. \\

\noindent\textbf{Implementations on other platforms:}
Although we implemented Medha in Alveo U250, the architecture is not tied to the specific FPGA. Large Xilinx FPGAs use the SSI technology~\cite{xilinx_ug872} to combine multiple Super Logic Regions (SLR). Therefore, the proposed ring of RPAUs (Sec.~\ref{sec:rpau_ring}) will be essential to implementing multi-RPAU HE architecture on large Xilinx FPGAs.   
Furthermore, the proposed ring interconnection of RPAUs can be extended to multi-FPGA implementations. Multi-FPGA platforms e.g., Amazon AWS EC2 F1 instances~\cite{amazon_aws} or~\cite{multi_fpga_s2c} keep the FPGAs in ring formation where each FPGA communicates with its neighbouring FPGAs using high-bandwidth bus (e.g., 400 Gbps bidirectional in~\cite{amazon_aws}). Therefore, during the key-switching operation, residue polynomials can be broadcast in low latency over the ring similar to the RPAU-to-RPAU broadcast.     
Multi-FPGA implementation of Medha will enhance its performance further and enable implementations for larger parameter sets. Each FPGA will host a set of RPAUs. Bootstrapping will definitely need several large FPGAs due to its sheer computational cost. 
In comparison to large single-chip accelerator architectures such as~\cite{feldmann_2021f1,BTS_isca,CraterLake,ARK}, multi-FPGA architectures will be modular by nature, making designing, prototyping and testing convenient while keeping re-usability. Furthermore, compared to the multi-million dollar ASIC architectures, a multi-FPGA platform will significantly cheaper. One VU13P FPGA (which is present inside U250 card) costs less than 9K US\$, and hence using eight of them will still be at least 50 times cheaper than fabricating~\cite{feldmann_2021f1} in silicon. Furthermore, FPGAs can be rented on hourly basis from the AWS. 

Medha can be ported to ASIC technologies. Medha has been described using Verilog RTL and only the BRAMs/URAMs and DSP multipliers are Xilinx IPs. In ASIC, BRAMs and URAMs will be replaced by dual-port SRAMs with the same functional and timing behavior. Similarly, the DSP multipliers will be replaced by normal multipliers.  
Interesting, the proposed ring interconnection of RPAUs will be ideal for an ASIC implementation. In the ring, neighboring RPAUs will be placed side by side on the layout of the chip. We anticipate that Medha's clock frequency will improve by three to five times depending on the ASIC technology.
One RPAU of Medha is synthesized in 65nm standard cell ASIC library and it meets 600 MHz clock with 4.6 $mm^2$ area. With newer technology, e.g., 12nm ASIC, we anticipate that ~1 GHz clock and a smaller area will be achievable. \\

\noindent\textbf{Verification of Medha:}
We verified the functional correctness of Medha in a real FPGA. Different homomorphic procedures, namely addition, multiplication, relinearization, rescaling, key-switching, rotation, etc., are computed correctly by Medha. All homomorphic procedures have branch-less and fixed control flow, making their testing convenient.

We consider formal verification of Medha as a future work. Complementing functionality testing, model checking-based formal verification could be used to check if the architecture description (e.g., HDL code, netlist, etc.) matched the specification. If performed correctly, formal verification could detect corner case bugs that may remain undetected using testing. Model checking-based formal verification assumes that the golden reference model is accurate. Note that any mistake in the model description could give a false confidence that the design is correct. Developing a complete and accurate model of a large HE accelerator (that includes hundreds or thousands of parallel cores, different types of memory elements, asynchronous IP blocks, etc.) will be a very challenging and time-taking research topic. Besides, formal verification is not advanced enough to take physical properties (e.g., timing violations) of the implementation platform into account. 
An emulated or simulated model of a hardware is not comparable with a working ASIC/FPGA prototype. Therefore, prototype testing cannot be replaced by formal verification of an architecture description. Neither prototyping nor formal verification is used in~\cite{feldmann_2021f1,BTS_isca,CraterLake} to bring confidence on the functional correctness of these massively large FHE architectures. Only~\cite{basalisc} reports formal verification of internal functional units, although no actual prototype is developed yet.\\

\noindent\textbf{Comparisons with hardware accelerators in other fields of research:}
Comparing our hardware acceleration results with hardware acceleration results presented in other fields of research will help us get a bigger picture. Hardware acceleration of AI is a very popular research topic with many researchers involved. 
Acceleration of neural networks on FPGA platforms show 15$\times$ to 150$\times$ speedups~\cite{ai_fpga1, ai_fpga2}. In general, ASIC implementations show up to 10x performance improvements compared to FPGAs due to higher clock frequency and additional customization. Intel’s latest AI chip~\cite{intel_ai_chip} obtains up to 1,000$\times$ speed up compared to CPUs. 
Our Medha in FPGA shows 134$\times$ faster latency than an Intel CPU running at 1.8 GHz. We anticipate that another 5$\times$ to 10$\times$ speedup will be feasible if an ASIC-optimized instance of Medha is fabricated.

\section{Conclusion}\label{sec:conc}

Despite being theoretically sound, HE suffers from performance issues due to its massive computational costs. Hence, hardware accelerators are crucial for making HE fast and practical. Also, providing support for multiple scheme parameters is another important requirement for hardware accelerator as different applications demand different scheme parameters.
%
In this paper, we first propose a flexible design methodology for the polynomial degree. It enables supporting multiple polynomial degrees using a fixed hardware architecture.
Then, we propose Medha, a flexible and programmable accelerator architecture that has been designed pragmatically to overcome the speed and flexibility limitations of previous HE accelerators. 
Medha gains efficiency from its highly optimized polynomial arithmetic blocks, parallel processing of instructions inside RPAU, parallel processing of residue polynomials using many RPAUs, highly efficient ring-based interconnection of RPAUs for data exchange, and customized on-chip memory unit. 
The memory-conservative design approach used to build Medha made it possible for the first time to compute key-switching without requiring any off-chip communication even for large HE parameters. 
Medha also utilizes the proposed flexible design methodology for the polynomial degree and it supports polynomial degrees $2^{14}$ and $2^{15}$ with coefficient modulus up to 546 bits.
Medha presents the first FPGA-based hardware accelerator for the RNS-HEAAN scheme with the polynomial degree $2^{15}$.

Medha was implemented in the Xilinx Alveo U250 card and accurate performance results were obtained for two large parameter sets. %
Compared to the highly-optimized SEAL~\cite{SEAL3.6} library, our Medha achieved up to $68\times$ and $78\times$ speedup on an Intel i5 CPU for the parameters $(\log Q, N) = (438, 2^{14})$ and $(546, 2^{15})$, respectively.
Medha achieved almost $2.37\times$ latency improvement compared to the block-pipelined and specific hardware accelerator HEAX~\cite{heax} for the parameter $(\log Q, N) = (438, 2^{14})$. The speed improvement shows that hardware accelerators for HE can attain high performance without losing programmability or flexibility.    